\documentclass[10pt]{report}

\usepackage{chapeqn}
\usepackage{epsf}
\usepackage{amsmath}
\usepackage{amsfonts}
\usepackage{amssymb}
\usepackage{bm}
\usepackage{bbm}
\usepackage{graphicx}
\usepackage{epsfig}
\usepackage{latexsym}
\usepackage{nicefrac}
\usepackage{color}
\usepackage{cmbright}
\usepackage[latin2]{inputenc}

\def\corresponds{{\lower.1.377ex\hbox{=}}{\rm\kern-.75em^\triangle}}
\def\succsim{\succ\kern-.9em_\sim\kern.3em}
\def\precsim{\prec\kern-1em_\sim\kern.3em}
\def\slantfrac#1#2{\kern1em^{#1}\kern-.3em/\kern-.1em_{#2}}
\def\lfrac#1#2{{}^{#1\!}\kern-.0em/_{#2}}
\def\trunc#1{[\mkern - 2.5 mu  [#1] \mkern - 2.5 mu ]}
\def\buildrel#1\under#2{\mathrel{\mathop{\kern0pt #2}\limits_{#1}}}

\def\tr{{\rm Tr}\,}

\definecolor{light}{gray}{0.90}
\definecolor{darker}{gray}{0.50}
\definecolor{dark}{gray}{0.30}

\usepackage{fancyhdr}
\renewcommand{\chaptermark}[1]%
{\markboth{\MakeUppercase{\chaptername~\thechapter: #1\\[2ex]}}{}}
\renewcommand{\chaptermark}[1]{\markright{\thechapter~#1}}

\def\uu{{\underline{u}}}

\def\dd{{\mathrm{d}}}
\def\ii{{\mathrm{i}}}
\def\ee{{\mathrm{e}}}

\def\calC{{\mathcal{C}}}
\def\calO{{\mathcal{O}}}
\def\calS{{\mbox{\sf{S}}}}
\def\cals{{\mbox{\sf{s}}}}

\def\uu{{u}}
\def\UU{{\mathcal{U}}}
\def\VV{{\mathcal{V}}}
\def\WW{{\mathcal{W}}}
\def\CC{{\mathcal{C}}}
\def\TT{{\mathcal{T}}}
\def\SW{{\mathrm{S}}}

\def\GG{{G}}
\def\tGG{\widetilde{G}}

\def\half{   {\textstyle{\frac12}}   }
\def\third{   {\textstyle{\frac13}}   }

\def\rept{{\mathrm{Re}}}
\def\impt{{\mathrm{Im}}}
\def\tfrac#1#2{ {\textstyle{\frac{#1}{#2}} } }
\def\trunc#1{[\mkern - 2.5 mu  [#1] \mkern - 2.5 mu ]}

\newcommand{\PT}{${\mathcal{PT}}$}

\DeclareSymbolFont{operators}   {OT1}{cmr} {m}{n}
\DeclareSymbolFont{letters}     {OML}{cmm} {m}{it}
\DeclareSymbolFont{symbols}     {OMS}{cmsy}{m}{n}
\DeclareSymbolFont{largesymbols}{OMX}{cmex}{m}{n}
\SetSymbolFont{operators}{bold}{OT1}{cmr} {bx}{n}
\SetSymbolFont{letters}  {bold}{OML}{cmm} {b}{it}
\SetSymbolFont{symbols}  {bold}{OMS}{cmsy}{b}{n}
\DeclareSymbolFontAlphabet{\mathrm}    {operators}
\DeclareSymbolFontAlphabet{\mathnormal}{letters}
\DeclareSymbolFontAlphabet{\mathcal}   {symbols}
\DeclareMathAlphabet      {\mathbf}{OT1}{cmr}{bx}{n}
\DeclareMathAlphabet      {\mathsf}{OT1}{cmss}{m}{n}
\DeclareMathAlphabet      {\mathit}{OT1}{cmr}{m}{it}
\DeclareMathAlphabet      {\mathtt}{OT1}{cmtt}{m}{n}
\SetMathAlphabet\mathsf{bold}{OT1}{cmss}{bx}{n}
\SetMathAlphabet\mathit{bold}{OT1}{cmr}{bx}{it}

\begin{document}

\pagestyle{empty}

{

\Large

\vspace*{1.0in}

\color{red}
\flushright{\Large \bf Multi--Instantons and Exact Results III:}
\flushright{\Large \bf Unification of Even and Odd Anharmonic Oscillators}

\vspace{1.5in}

\color{blue}
\large
\flushright{Ulrich D. Jentschura\\
{\em Department of Physics}\\
{\em Missouri University of Science and Technology}\\
{\em Rolla, Missouri, MO65409, USA}\\[2ex]
Andrey Surzhykov}\\
{\em Physikalisches Institut der Universit\"{a}t}\\
{\em Philosophenweg 12, 69120 Heidelberg, Germany and}\\
{\em GSI, 64291 Darmstadt, Germany}\\[2ex]
Jean Zinn--Justin\\
{\em CEA, IRFU and Institut de Physique}\\
{\em Th\'{e}orique, Centre de Saclay}\\
{\em F-91191 Gif-Sur-Yvette, France}\\
}

\color{black}

\vfill

\pagestyle{empty}

\newpage

\vspace*{0.0cm}
\begin{center}
\begin{tabular}{c}
\hline
\rule[-3mm]{0mm}{12mm}
{\large \sf Multi--Instantons and Exact Results III:}\\
\rule[-5mm]{0mm}{12mm}
{\large \sf Unification of Even and Odd Anharmonic Oscillators}\\
\hline
\end{tabular}
\end{center}
\vspace{0.0cm}
\begin{center}
Ulrich D. Jentschura\\
\vspace{0.2cm}
\scriptsize
{\em
Department of Physics,
Missouri University of Science and Technology,\\
Rolla, Missouri, MO65409, USA}
\end{center}

\begin{center}
Andrey Surzhykov\\
\vspace{0.2cm}
\scriptsize
{\em Physikalisches Institut der Universit\"{a}t,
Philosophenweg 12, 69120 Heidelberg, Germany and\\
GSI Helmholtzzentrum f\"ur Schwerionenforschung, 64291 Darmstadt,
Germany}
\end{center}

\begin{center}
Jean Zinn--Justin\\
\vspace{0.2cm}
\scriptsize
{\em CEA, IRFU and Institut de Physique Th\'{e}orique,}\\
{\em Centre de Saclay, F-91191 Gif-Sur-Yvette, France}
\end{center}
\vspace{0.3cm}
\begin{center}
\begin{minipage}{14.0cm}
{\underline{Abstract}}
This is the third article in a series of three papers on
the resonance energy levels of anharmonic oscillators.
Whereas the first two papers mainly dealt with
double-well potentials and modifications thereof
[see J.~Zinn-Justin and U.~D.~Jentschura,
Ann.~Phys.~(N.Y.) {\bf 313} (2004), pp.~197 and 269],
we here focus on simple even and odd anharmonic oscillators
for arbitrary magnitude and complex phase of the coupling parameter.
A unification
is achieved by the use of \PT-symmetry inspired dispersion
relations and generalized quantization conditions that include
instanton configurations.  Higher-order formulas
are provided for the oscillators of degrees $3$ to $8$,
which lead to subleading corrections to the
leading factorial growth of the perturbative coefficients
describing the resonance energies.
Numerical results are provided, and higher-order terms
are found to be numerically significant.
The resonances are described
by generalized expansions involving
intertwined non-analytic exponentials, logarithmic terms
and power series. Finally, we
summarize spectral properties and dispersion
relations of anharmonic oscillators, and their interconnections.
The purpose is to look at one of the
classic problems of quantum theory from a new perspective,
through which we gain systematic access to the phenomenologically significant
higher-order terms.
\end{minipage}
\end{center}

\vspace{0.6cm}

Asymptotic problems and properties;\\
Summation of perturbation theory\\
\vfill

\newpage

\tableofcontents

\newpage

\clearpage\fancyhead[R]{\normalsize \rightmark}
\pagestyle{fancy}

% \tableofcontents

%
% Introduction
%
\chapter{Introduction}

\vspace*{-0.4cm}
\textcolor{light}{ \rule{\textwidth}{0.2cm} }

We here continue our investigations on anharmonic oscillators,
shifting the focus away from double-well and related potentials,
which had been the subject of the first two
papers~\cite{ZJJe2004i,ZJJe2004ii} of this series. Following ideas
outlined previously in a brief exposition~\cite{JeSuZJ2009prl},
we here elaborate on a unified theory of even and odd anharmonic
oscillators, where the resonance energies are determined by a
harmonic oscillator Hamiltonian which is perturbed by an even or odd
monomial of the coordinate.

Indeed, anharmonic oscillators represent a very well developed field
of study, and the scientific literature on the subject is extensive
(for a few exemplary literature references,
see~\cite{GrGrSi1970,BaCo1971,BeWu1973,Vo1983,We1996b,We1996d}).
Still, it is quite surprising that a few basic questions
regarding the anharmonic, one-dimensional oscillators
have apparently not yet been fully addressed
to the best of our knowledge.
Some of these unanswered questions concern the general
structure of resonance energies of the anharmonic oscillators;
the imaginary part of a resonance energy
is nonvanishing and describes the tunneling of the
bound particle through the potential well.
We recall that resonance energies with a nonvanishing imaginary part are
relevant, e.g., for even anharmonic oscillators when the coupling parameter $g$
in the interaction $g\, x^N$ becomes negative ($N$ even), and for odd
anharmonic oscillators when the coupling parameter $g$ in the interaction
$\sqrt{g}\, x^M$ is positive ($M$ is assumed to be an
odd integer, the motivation for the
square root is explained below).  Here, we attempt to unify the
treatment of the resonance eigenvalues of anharmonic oscillators of
even and odd degree, with a special emphasis on the dispersion relations
fulfilled by the resonance energies.

We also investigate higher-order formulas
(``hyperasymptotics'') for the oscillators of arbitrary degree. These have the
form of generalized, nonanalytic expansions, where a nonanalytic (in the
coupling constant) prefactor is multiplied by logarithmic and power series
(this constitutes a triple expansion).
Note that generalized nonanalytic expansions, which contain
expressions like $\exp(-A/g^b)$ with positive constants $A$ and $b$,
logarithmic terms like $\ln(g)$, and powers of $g$, have recently been shown to
be relevant to the description of energy levels of a number of one-dimensional
quantum systems~\cite{ZJJe2004i,ZJJe2004ii}, including problematic cases where
the ground-state energy vanishes to all orders of perturbation theory, but the
true ground-state energy is manifestly different from zero and dominated by a
nonanalytic factor of the form $\exp[-1/(3g)]$ (see
Refs.~\cite{HeSi1978plb,CaGrMa1988,CaGrMa1996,JeZJ2004plb}).
Anharmonic
oscillators, as shown here, also imply triple expansions for a complete
description of the resonance energies.  Because of their
phenomenological importance, the anharmonic oscillators occupy a special
position within quantum theory.

Let us therefore summarize some of the most prominent questions
investigated in the current report:
How can we write down a satisfactory
expansion, in powers of the coupling, describing the
unstable cases where the coupling parameter multiplying the
anharmonic perturbation attains a value that
makes the levels unstable?
Do resonance eigenvalues persist when we increase the
modulus of the coupling parameter to infinity?
What is the general behaviour of large-order
perturbation theory, for arbitrary resonance energies
of odd anharmonic oscillators of arbitrary degree?

Although the above questions have not yet been
addressed in the literature to the best of our knowledge,
many physical aspects of anharmonic oscillators are indeed well known.
E.g., the spectrum of an
even anharmonic oscillator (with a perturbation of
the form $g\, x^N$ with $g > 0$ and even
integer $N$) is bounded from below and
consists of discrete energy eigenvalues,
when the Hamiltonian is endowed with $L^2(\mathbbm{R})$ boundary conditions.
The divergent perturbation series describing the
energy eigenvalues of even oscillators are
known to be Borel summable~\cite{GrGrSi1970}.
In addition, one may derive effective
semi-analytic approximations to these energies
such as renormalized strong-coupling
expansions~\cite{We1996b,We1996d} in the ``stable'' sector
of positive coupling parameter $g > 0$, which are
applicable to both weak and strong coupling. However,
when $g$ is continued into the complex plane
and the boundary conditions are smoothly deformed in the
sense of complex scaling~\cite{BaCo1971},
then, as shown in Ref.~\cite{BeWu1973},
the resonance energy eigenvalues become complex. Indeed,
the smooth functions $E_n(N, g)$ describing the $n$th
resonance energy eigenvalue of an even oscillator
of degree $N$ are analytic functions of the (complex)
coupling $g$ for $-\pi < {\rm arg}\,g < \pi$.
There are branch cuts at negative $g$, and the
resonance energies are of the form ${\rm Re}\, E_n(N,g)
\pm {\rm i}\, {\rm Im}\, E_n(N,g)$, where the two different signs are
attained on different sides of the branch cut along the negative axis.

For odd oscillators, by contrast, the situation is not
as straightforward and has been discussed in the
literature quite recently~\cite{BeDu1999}.
The most straightforward choice for the interaction term, namely $g\,x^M$,
does not lead to a satisfactory description
because the transformation $g\to -g$ in this
case is equivalent to a parity transformation which
changes the Hamiltonian but leaves its spectrum invariant.
It may seem counterintuitive that the most natural
definition of the perturbative term in this case involves
the structure $\sqrt{g}\,x^M$.
In this case, employing the concept of
\PT-symmetry~\cite{BeBo1998,BeBoMe1999,%
BeWe2001,BeDuMeSi2001,BeBrJo2002},
one can show that the spectrum for $L^2(\mathbbm{R})$-boundary
conditions is real, discrete and
bounded from below for negative coupling $g < 0$.
Note that the case of negative coupling
corresponds to a \PT-symmetric interaction term
$\pm \ii \, \sqrt{|g|} \, x^M$, where again the $\pm$ sign of the
perturbative term does not matter as far as the spectral
properties are concerned. Starting from the case of
negative $g$, one can then smoothly deform the boundary
conditions when varying $g$ in the complex plane,
so that a branch cut for the energy emerges for positive coupling.
Resonances (with a negative imaginary part of the energy)
or antiresonances (with a positive imaginary part of the energy)
appear infinitesimally displaced above and below
the positive real $g$-axis, respectively.

It has been known for a long time that the
perturbation series describing the energy levels
of anharmonic oscillators constitute asymptotic, divergent series.
In 1971 (see Ref.~\cite{BeWu1971}), the
general leading asymptotics for the perturbative coefficients
describing the energy of an arbitrary excited level
of an even oscillator of arbitrary degree were derived.
Here, we generalize this result to the case of odd
oscillators of arbitrary degree, and for an
arbitrary excited level. This generalization is made possible
by a combination of
(i) \PT-symmetry-based dispersion relations derived
in Ref.~\cite{BeDu1999}, and (ii) by generalized quantization
condition that are conjectured in the current work.

While the unification of even and odd anharmonic oscillators
necessarily cannot lead to completely analogous
formulas for both cases, our conjectured
unified quantization conditions cover a number
of cases simultaneously: (i) real and positive coupling
for even and the ${\mathcal PT}$-symmetric case for odd
oscillators, and (ii) negative real coupling for even
oscillators and positive real coupling for
odd anharmonic oscillators.
A ``master table'' with some currently known, important properties
of oscillators of both parities is also provided
(see below in Sec.~\ref{master}).
We also emphasize the subleading corrections to
various physical quantities of interest, such as decay
rates of unstable resonances. These are provided for the
anharmonic oscillators of degree $3$ (cubic), $4$ (quartic),
$5$ (quintic), $6$ (sextic), $7$ (septic) and $8$ (octic
oscillator).

The \PT-symmetry of the odd oscillators, for negative
coupling, is used here as a tool facilitating the formulation of
the problem leading to the general large-order asymptotics of the
perturbation theory characterizing the odd oscillators.
Note that \PT-symmetric Hamiltonians, which represent
a natural generalization of Hermitian Hamiltonians
when one allows for a non-standard scalar product
of Hilbert-space vectors~\cite{BeBo1998,Mo2002i,Mo2002ii,Mo2002iii,Mo2005cubic},
have recently found applications in many areas of physics.
Examples include bound-state scattering
theory~\cite{Ma2006,Zn2008},
physical realizations of ${\mathcal{PT}}$-symmetric potentials
in optical waveguides~\cite{RuDeMu2005,MoLo2008,MuMaEGCh2008,KrTa2008},
and the description of squeezed states of atoms~\cite{DeNaDR2005},
as well as astrophysical contexts~\cite{AnCaGiKaRe2008}.
Recently, the Lee model~\cite{BeBrChWa2005} and the
Pais--Uhlenbeck oscillator model~\cite{BeMa2008} have been
shown to be free from problematic ghost states with negative norm,
provided one endows the underlying Hilbert space
with a non-standard scalar product that ensures the
equivalence of the model with a Hermitian theory and
makes the time evolution unitary.
Here, \PT{}-symmetry is applied to one of
the classic problems of quantum theory.

The general structure of our investigations is as follows.
A short digression of the basic formulas
underlying our investigation, and a brief discussion
of the derivation of the generalized quantization conditions
is provided in Sec.~\ref{basic}.
The leading-order general formulas for odd oscillators
are also discussed in Sec.~\ref{basic}.
We then proceed to the discussion of higher-order formulas
(``hyperasymptotics'') in Sec.~\ref{higher}.
The anharmonic oscillators of the third up to the
eighth degree are treated in
Secs.~\ref{cubic}---\ref{octic}, respectively.
A ``master table'' displaying some properties of even versus odd anharmonic
oscillators is provided in Sec.~\ref{master}.
Numerical calculations for weak (see Sec.~\ref{numdiag})
and strong coupling (see Sec.~\ref{numlarge}) verify the
analytic results.
Conclusions are reserved for Sec.~\ref{conclu}.

%
% From Instantons and Dispersion Relations to General Formulas
%
\chapter{From Instantons and Dispersion Relations to General Formulas}
\label{approach1}

\vspace*{-0.4cm}
\textcolor{light}{ \rule{\textwidth}{0.2cm} }

\section{Definition of the Hamiltonians}
\label{defham}

Before we indulge into the heart of the investigations,
let us first mention a few conventions that are
used throughout  this article. The degree of an oscillator
is denoted by $m$ for a perturbation proportional to $q^m$.
Unless stated otherwise, we replace $m \to N$ if $m$ is even,
and by $m \to M$ if $m$ is odd. The quantum number of the $n$th
excited level of a one-dimensional
oscillator is denoted by $n = 0,1,2,\ldots$.
We denote the Hamiltonian of an even oscillator by
$H_N(g)$, where $g$ is the coupling,
\begin{subequations}
\begin{equation}
\label{HN}
H_N(g) =
-\frac12 \, \frac{\partial^2}{\partial q^2} + \frac12 \, q^2 +
g\; q^N \,.
\end{equation}
The resonance energies of the even oscillator
are $E^{(N)}_n(g)$. For the Hamiltonian $h_M(g)$ of an odd oscillator,
we use the convention
\begin{equation}
\label{hM}
h_M(g) =
-\frac12 \, \frac{\partial^2}{\partial q^2} + \frac12 \, q^2 +
\sqrt{g}\,\, q^M \,,
\end{equation}
\end{subequations}
with resonance energies $\epsilon^{(M)}_n(g)$. There are two reasons
for the appearance of the square root of the coupling parameter in
Eq.~\eqref{hM}. The first reason is that within perturbation theory,
the square root ensures that the perturbation series for the
resonance energies of the odd contain only integer powers of the
coupling, and that the first nonvanishing perturbation is of order
$g$, not $g^2$. The second reason is connected to the analytic
structure of the dispersion relations for the resonance energies, as
discussed in Sec.~\ref{dispersion} below. Roughly speaking, the
square root is necessary in order to ensure a one-to-one mapping of
the spectrum of resonances of~\eqref{hM} with a particular value of
$g$. For purely imaginary coupling $\sqrt{g} = \ii\,\beta$, with
real $\beta$, the spectrum of $h_M(g)$ is real by virtue of
\PT{}-symmetry. Furthermore, the spectrum is the same for $\sqrt{g}
= \ii\,\beta$ and $\sqrt{g} = -\ii\,\beta$, as well as the
corresponding eigenfunctions (up to a parity transformation). Both
of the mentioned values of $\sqrt{g}$ are uniquely associated with
the value $g = -\beta^2$, and it is this one-to-one mapping of the
resonance eigenvalue to a value of $g$ that we need in order to
formulate a dispersion relation for the resonance energies, which is
discussed below.

\begin{figure}[tb!]
\begin{center}
\begin{minipage}{0.9\linewidth}
\begin{center}
\parbox{0.35\linewidth}{
\includegraphics[width=1.0\linewidth]{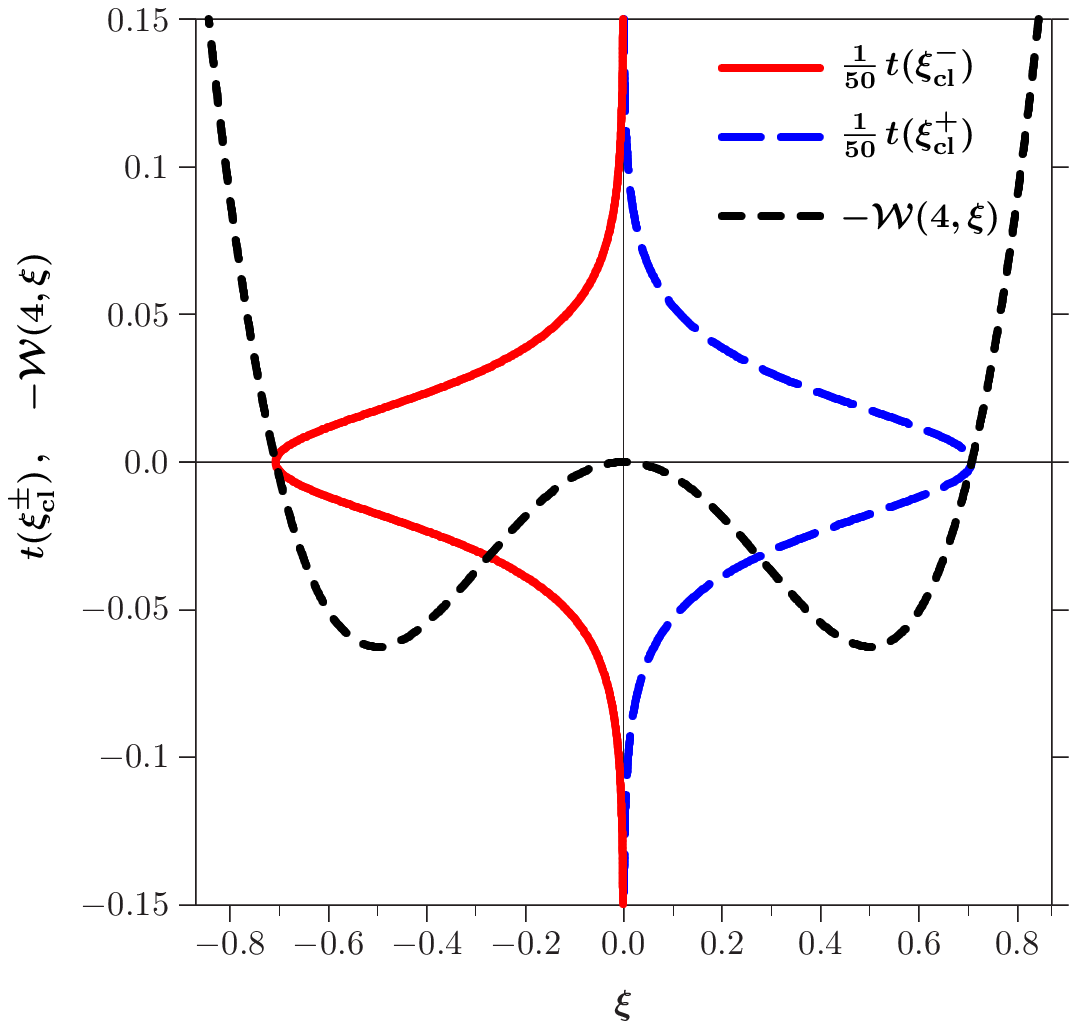} \\
\centerline{$\qquad$ (a)}}
\parbox{0.35\linewidth}{
\includegraphics[width=1.0\linewidth]{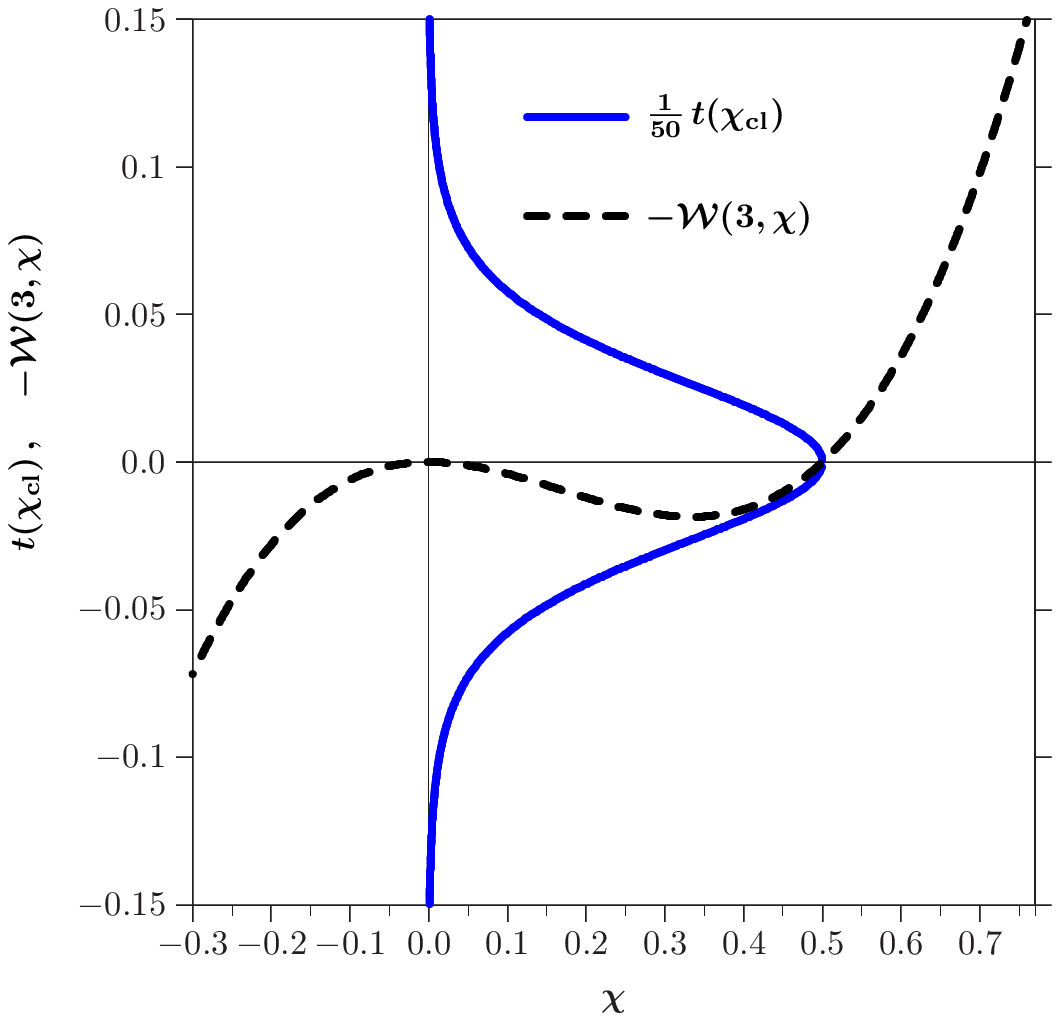} \\
\centerline{$\qquad$ (b)}}\\[5ex]
\parbox{0.35\linewidth}{
\includegraphics[width=1.0\linewidth]{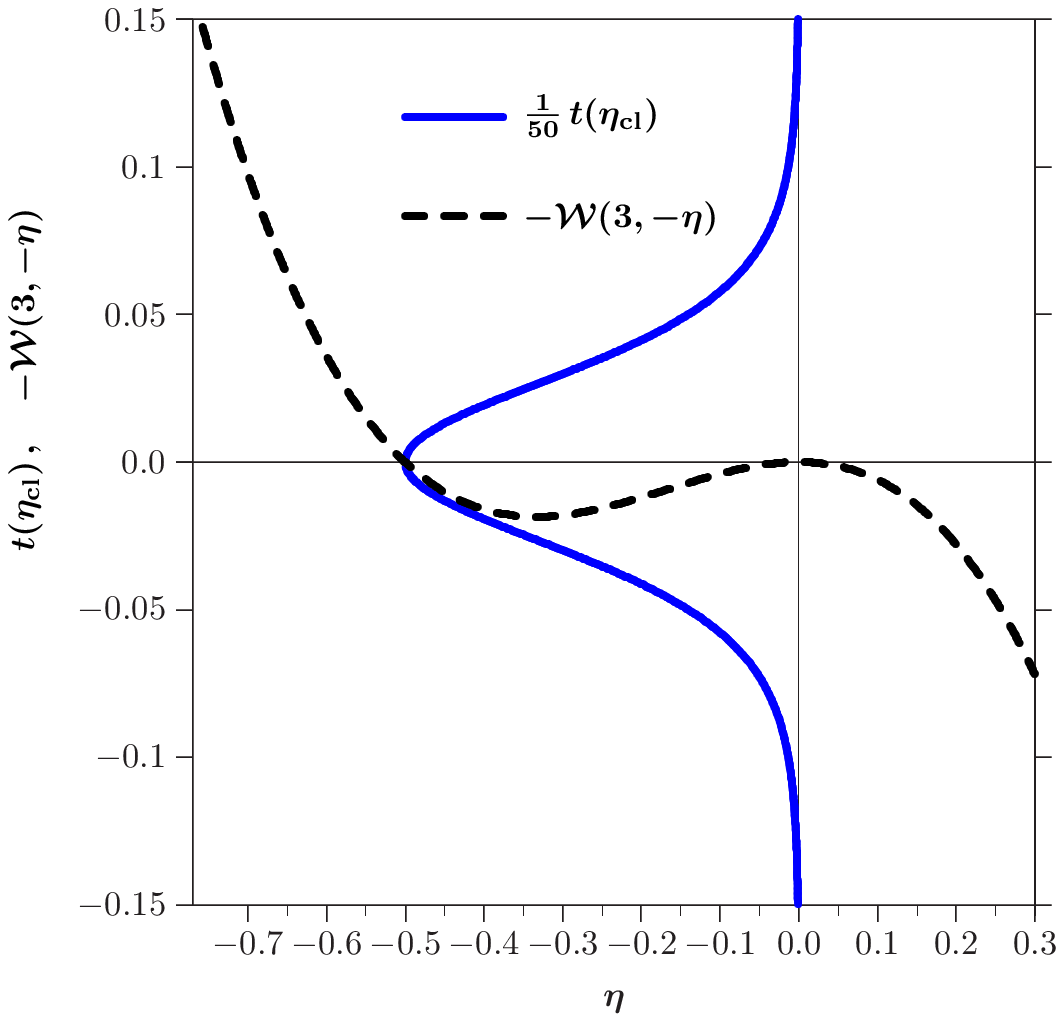} \\
\centerline{$\qquad$ (c)}}
\parbox{0.35\linewidth}{
\includegraphics[width=1.0\linewidth]{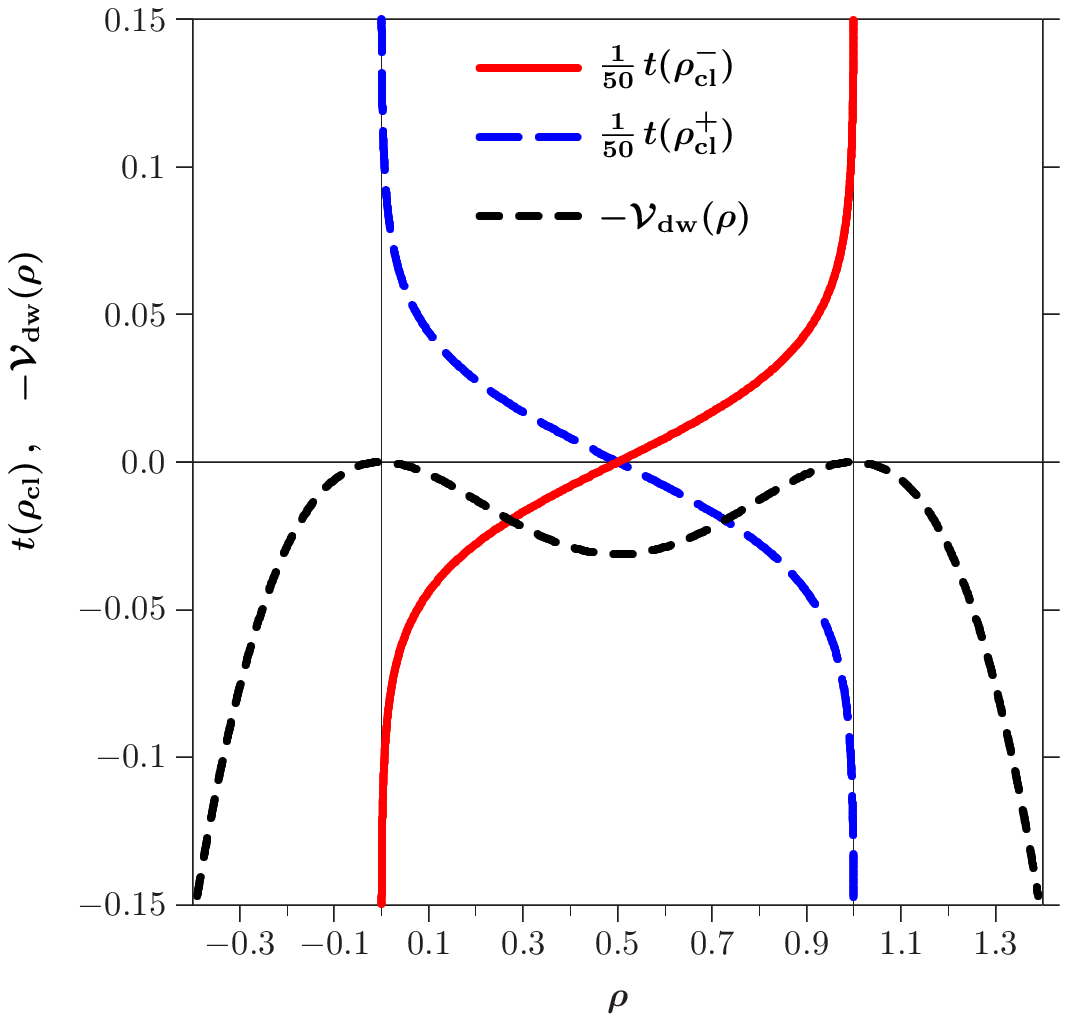} \\
\centerline{$\qquad$ (d)}}\\[5ex]
\caption{\label{fig1} Trajectories (world lines, the time axis is the ordinate)
of several instantons. The quartic ($N=4$)
instanton configuration $\chi^\pm_{\rm cl}(t)$ is given in
Eq.~(\ref{chiclpm}) and shown in Fig.~(a).
The cubic ($M=3$) instanton $\chi_{\rm cl}(t)$ is
given in Eq.~(\ref{chicl}) and shown in Fig.~(b).
The cubic instanton $\eta_{\rm cl}(t) = -\chi_{\rm cl}(t)$
is found for the reflected cubic potential, where the sign of the
$\eta^3$ term is inverted [see Eq.~\eqref{etacl}].
Figure~(d) shows the $\rho^\pm_{\rm cl}$ instanton which
exists for the double-well potential, according to Eq.~\eqref{rhocl}.}
\end{center}
\end{minipage}
\end{center}
\end{figure}

%
% Instanton Trajectories
%
\section{Instanton Trajectories}

Instantons are nontrivial saddle points of the Euclidean
action of a quantum theory, as explained in the
first two papers of the current series~\cite{ZJJe2004i,ZJJe2004ii}.
The Euclidean action
is the action obtained by varying the total energy
(kinetic plus potential energy) of a particle
along a classical path. The partition function of the
anharmonic oscillator can be written as
\begin{equation}
\label{Zbeta}
Z(\beta) = \tr \exp(-\beta H) \,,
\end{equation}
where $H$ is the Hamiltonian of the system
[$H = H_N(g)$ for an even oscillator, and
$H = h_M(g)$ for an odd oscillator; in our notation,
we suppress the dependence of the right-hand side on the
coupling parameter $g$].
A path integral representation of $Z(\beta)$ involves
periodic classical orbits,
\begin{equation}
\label{ZbetaPath}
{\rm Tr} \, \exp( - \beta \, H ) =
\int \dd q_0
\int\limits_{q\left(-\frac{\beta}{2}\right) =
q\left(\frac{\beta}{2}\right) = q_0}
[{\rm d}q(t)] \, \exp \left\{
-\calS[-\tfrac12 \beta, \tfrac12 \beta, q] \right\} \,,
\end{equation}
where the integration over the end point $q_0$ of the
classical trajectory (which is equivalent to its starting point)
is explicitly given. The Euclidean action
$\calS[-\tfrac12 \beta, \tfrac12 \beta, q]$ is the
integral over the kinetic and potential energy
of the system as it evolves from Euclidean time $-\tfrac12 \beta$
to $\tfrac12 \beta$. For an even oscillator of degree $N$, we have
\begin{equation}
\label{SN}
\calS_N[ -\tfrac12 \beta, \tfrac12 \beta, q] =
\int\limits_{-\beta/2}^{\beta/2} {\rm d}t \, \left[ \frac{1}{2} \,
\left( \dot{q}(t)^2 + q(t)^2 \right) + g\, q(t)^N \right]  \,,
\end{equation}
whereas for an odd oscillator of degree $M$,
\begin{equation}
\label{sM} \cals_M[ -\tfrac12 \beta, \tfrac12 \beta, q] =
\int\limits_{-\beta/2}^{\beta/2} {\rm d}t \, \left[ \frac{1}{2} \,
\left( \dot{q}(t)^2 + q(t)^2 \right) + \sqrt{g}\, q(t)^M
\right] \,.
\end{equation}
In order to calculate the perturbative expansion of, say,
the ground-state energy of an anharmonic oscillator,
one expands about the trivial saddle point of the
Euclidean action, which is given by the classical trajectory
$q(t) = 0$. One writes
\begin{subequations}
\label{quantum_fluc}
\begin{align}
\calS_N[ -\tfrac12 \beta, \tfrac12 \beta, q] =& \;
\frac12
\int\limits_{-\beta/2}^{\beta/2} {\rm d}t
\int\limits_{-\beta/2}^{\beta/2} {\rm d}t'
q(t) \, M(t, t') \, q(t')
+ g  \int\limits_{-\beta/2}^{\beta/2} {\rm d}t \, q(t)^N \,,
\\[0.3ex]
\cals_M[ -\tfrac12 \beta, \tfrac12 \beta, q] =& \;
\frac12
\int\limits_{-\beta/2}^{\beta/2} {\rm d}t
\int\limits_{-\beta/2}^{\beta/2} {\rm d}t'
q(t) \, M(t, t') \, q(t')
+ \sqrt{g} \int\limits_{-\beta/2}^{\beta/2} {\rm d}t \, q(t)^M \,,
\\[0.3ex]
M(t, t') =& \; \delta(t - t') \,
\left( - \frac{\partial^2}{\partial t'^2} + 1 \right) \,.
\end{align}
\end{subequations}
The inverse of $M$ is the propagator
\begin{equation}
\label{defDelta}
\Delta(t, t') = \int\limits_{-\infty}^{\infty} \frac{{\rm d} p}{2 \pi}\,
\frac{{\rm e}^{{\rm i}\, p\, (t-t')}}{p^2 + 1} =
\frac{1}{2} \, \exp( - | t - t' |) \,.
\end{equation}
As described in Ref.~\cite{ZJ2003},
the perturbative expansion for the ground-state energy
\begin{equation}
E_0(g) = \lim_{\beta \to \infty} \left( -\frac{1}{\beta} \,
\ln \left( {\rm Tr} \, \exp( - \beta \, H ) \right) \right)
\end{equation}
involves Feynman diagrams which can be expressed in terms
of integrals of the progagator~\eqref{defDelta}.
Because the exponentiated [see Eq.~\eqref{Zbeta}]
quadratic quantum fluctuation about the
saddle point given by the $M$ operator in Eq.~\eqref{quantum_fluc}
is equivalent to a Gaussian integration measure,
one calls the saddle point given by the classical path
$q(t) = 0$ the Gaussian saddle point.

In view of the obvious dominance of the Gaussian saddle point of the
Euclidean action for the
perturbative expansion, one might wonder if other nontrivial saddle
points exist, and if yes, what role they might play.
The Euclidean action corresponding to the path $[q(t) = 0]$ is zero.
Instantons are classical paths which lead to a finite Euclidean
action in the limit of infinite Euclidean time $\beta$.
They do not correspond to a physical motion of the
particle in the potential under investigation, but rather, to a
classical motion in an inverted potential.
We remember that the Lagrange function for a classical
particle is $T-V$, whereas in the Euclidean action we
are integrating over $T+V$ ($T$ is the kinetic energy, and
$V$ is the potential).

In order to appreciate the importance of the instanton
configurations (which are depicted in Fig.~\ref{fig1}),
we must consider a slightly more complicated situation,
namely, the cut of the partition function in the
complex plane which is present for ``unstable'' values of the
coupling parameter $g$, i.e., for those which lead to a potential
with unstable, relative minima and potential wells through
which the particle may tunnel~\cite{LGZJ1980,ZJ1996}. For the even oscillators as given
by the Hamiltonian~\eqref{HN}, the unstable coupling parameters
are given by $g < 0$, because the potential $\tfrac12 q^2 - |g|\, q^N$
approaches $-\infty$ for $q \to \pm \infty$.
For an odd oscillator [see Eq.~\eqref{hM}],
all positive values of $g$ lead to an unstable potential because
$\tfrac12 q^2 + g\, q^M \to -\infty$ for $q \to -\infty$.
The ground-state energy in this case acquires an imaginary part
and becomes a resonance energy.
The imaginary part of the resonance energy is related to the
imaginary part of the partition function according to
\begin{equation}
\impt E_0(g) =
\lim_{\beta \to \infty}
\left( -\frac{1}{\beta} \, \frac{ \impt \, Z(\beta) }{ Z(\beta) } \right) \,.
\end{equation}
The sign of the imaginary part of $E_0$ depends on the
sign of the (infinitesimal) imaginary part of $g$.
Discussing (for definiteness) the case of an even oscillator,
we define $E_0(g)$ to be the resonance energy [with $\impt E_0(g) < 0$]
for $g = -|g| + \ii 0$ and
to be the anti-resonance energy [with $\impt E_0(g) > 0$]
for $g = -|g| - \ii 0$.
The cut of the resonance energy then coincides with the
cut of the partition function.
It has been observed in Refs.~\cite{ZJ1996,ZJ2003} that the
contribution of the Gaussian saddle point to the
cut of the partition function, i.e., to
$\impt Z(\beta)$, vanishes, and that
$\impt Z(\beta)$ can be approximated very well by the instanton
configurations.

In order to find the instanton configuration for the even oscillator,
we scale the classical path as
$q(t) = (-g)^{-1/(N-2)} \, \xi(t)$ in Eq.~\eqref{SN}.
The Euclidean action becomes
\begin{equation}
\label{Seven}
\calS_N = (-g)^{-\frac{2}{N-2}} \,
\int \dd t \, \left( \half \, {\dot \xi}^2 + \half \, \xi^2 -
\xi^N \right) \,.
\end{equation}
The instanton
now is to be searched in the space of the paths $\xi(t)$,
whose Euclidean action remains finite when $t$ goes from $-\infty$ to $\infty$.
For $N= 4$
and $t_0 = 0 $, the motion corresponds to a classical trajectory in
the inverted potential $\xi^4 - \tfrac12 \xi^2$ and is depicted in
Fig.~\ref{fig1}(a). It becomes immediately clear that the instanton
configuration is twofold degenerate. Namely, starting for Euclidean
time $t \to \infty$ at a position $\xi(t = -\infty)$ infinitesimally
to the left or to the right of the central maximum of the inverted
potential, the particle may go through either one of the two
``troughs'' to the right or to left, before bouncing back toward the
central maximum at Euclidean time $t \to \infty$. According to
Refs.~\cite{BrLGZJ1977prd1,ZJ2003}, the corresponding solutions to
the (Euclidean) equations of motion read, for general $N$,
\begin{subequations}
\label{instantonEven}
\begin{align}
\label{qclpm}
q^{\pm}_{\rm cl}(t) = & \;
(-g)^{-\frac{2}{N-2}} \, \chi^\pm_{\rm cl}(t) \,,
\\[0.3ex]
\label{chiclpm}
\chi^\pm_{\rm cl}(t)
= & \; \pm \frac{1}{ \{ 1 + \cosh[(N-2) (t - t_0)] \}^{1/(N-2)}} \,.
\end{align}
\end{subequations}
Here, $t_0$ is the ``starting point'' of the instanton,
acting as a collective coordinate for the instanton configuration
from the point of view of the path integral formalism~\cite{ZJ2003}.
The instanton action is
\begin{align}
\label{instEven}
\calS_N[q^\pm_{\rm cl}] =& \; \left(- \frac{2}{g}\right)^{2/(N-2)} \,
B\left( \frac{N}{N-2}, \frac{N}{N-2} \right)
= \left(- \frac{1}{g}\right)^{2/(N-2)} \, {\cal A}(N) \,,
\\[1.377ex]
\label{definitionA}
{\cal A }(m) =& \; 2^{2/(m-2)} \,
B\left(  \frac{m}{m-2}, \frac{m}{m-2} \right) \,,
\end{align}
where $B(x, y) = \Gamma(x)\,\Gamma(y)/\Gamma(x + y)$ is the
Euler Beta function.

For odd anharmonic oscillators of degree $M$, we transform
$q(t) = - g^{-1/(2 M-4)} \chi(t)$ and obtain the
scaled Euclidean action [cf.~Eq.~\eqref{sM}]
\begin{equation}
\label{Sodd}
\cals_M =
g^{-\frac{1}{M-2}} \,
\int \dd t \, \left( \half \, {\dot \chi}^2 + \half \, \chi^2 -
\chi^M \right) \,.
\end{equation}
With this scaling,
the inverted potential now reads $\chi^M - \tfrac12 \chi^2$
and is of the same form as for the even oscillator.
The instanton trajectory $q(t) = q_{\rm cl}(t)$ is unique and reads
\begin{subequations}
\label{instantonOdd}
\begin{align}
\label{qcl}
q_{\rm cl}(t) = & \; g^{-\frac{1}{M-2}} \, \chi_{\rm cl}(t) \,,
\\[0.3ex]
\label{chicl}
\chi_{\rm cl}(t) = & \;
\frac{1}{ \{ 1 + \cosh[(M-2) (t-t_0)] \}^{1/(M-2)}} \,.
\end{align}
\end{subequations}
For the case $M = 3$ and $t = t_0$, the trajectory is shown in Fig.~\ref{fig1}(b).
Inserting $q_{\rm cl}(t)$ into Eq.~\eqref{Sodd},
we obtain the instanton action
\begin{align}
\label{instOdd}
\cals_M[q_{\rm cl}] =& \; \left(\frac{4}{g}\right)^{1/(M-2)} \,
B\left( \frac{M}{M-2}, \frac{M}{M-2} \right)
= \left(\frac{1}{g}\right)^{1/(M-2)} \, {\cal A}(M) \,.
\end{align}
Both instanton actions~\eqref{instEven} and~\eqref{instOdd}
are positive in the domain where the instantons exist
(for negative coupling in the case of an even oscillator and
for positive coupling in the case of an odd oscillator).

The formulas~\eqref{instEven} and~\eqref{instOdd} can
be unified. Namely, if we parameterize the perturbation as $g \, q^m$
for general (even or odd) $m$, then we can write the
instanton action as $(\mp 2/g)^{2/(m-2)} \,
B( \textstyle{ \frac{m}{m-2}, \frac{m}{m-2}})$, where the plus and minus signs
correspond to odd and even potentials, respectively.

The scaled instanton action ${\cal A}(m)$ is
expressible in terms of a scaled potential $\WW(m, q)$ where
the sign of the perturbative term has been inverted,
\begin{equation}
\label{defWm}
\WW(m, q) = \frac12 \, q^2 - q^m\,.
\end{equation}
As evident from Fig.~\ref{fig1},
the instanton configurations correspond to the motion of a classical particle in a
potential $-\WW(m, q)$. We obtain from the paths given in
(\ref{chiclpm}) and (\ref{chicl}),
\begin{equation}
\label{Am}
{\cal A}(m) =
\int\limits_0^{2^{-1/(m-2)}} \dd q \, \left( 2 \, \sqrt{2 \, \WW(m,q)} \right)
= 2^{2/(m-2)} \,
B\left( \frac{m}{m-2}, \frac{m}{m-2} \right) \,,
\end{equation}
where we note that $2^{-1/(m-2)}$ is a zero
of the potential $\WW(m,q)$.

The instanton path defined in Eq.~\eqref{qcl} fulfills
$q_{\rm cl}(t) > 0$. However, this instanton configuration
is not uniquely associated with the Hamiltonian~\eqref{hM}.
To see this, we start from an
odd anharmonic oscillators of degree $M$ and transform
$q(t) = g^{-1/(2 M-4)} \, \eta(t)$.
In this manner, we obtain the
scaled  Euclidean action [cf.~Eq.~\eqref{sM}]
\begin{equation}
\label{Sodd_eta}
\cals_M =
g^{-\frac{1}{M-2}} \,
\int \dd t \, \left( \half \, {\dot \eta}^2 + \half \, \eta^2 + \eta^M \right)\,.
\end{equation}
With this scaling, the inverted potential reads
$-\left( \eta^M + \tfrac12 \, \eta^2 \right) =
\left( -\eta\right)^M - \tfrac12 \, \eta^2$.
By comparison to Eq.~\eqref{chicl}, the instanton is found to be
\begin{equation}
\label{etacl}
\eta_{\rm cl}(t) =
-\frac{1}{ \{ 1 + \cosh[(M-2) (t-t_0)] \}^{1/(M-2)}} < 0\,.
\end{equation}
For $M = 3$ and $t_0 = 0$,
the corresponding instanton trajectory is shown in Fig.~\ref{fig1}(c).

For the double-well potential, the Euclidean action reads,
according to Eq.~(2.10a) of Ref.~\cite{ZJJe2004i},
\begin{equation}
\label{Sdw}
\calS_{\rm dw} =
\int \dd t \, \left( \frac{1}{2} \, {\dot \rho}^2 +
\frac{1}{2 g^2} \, \rho^2 (1 - \rho)^2 \right) \,.
\end{equation}
This expression contains the double-well potential
\begin{equation}
\label{Vdw}
{\cal V}_{\rm dw}(q) = \frac{1}{2} \, \rho^2 (1 - \rho)^2 \,.
\end{equation}
One of the possible instanton configurations is
\begin{subequations}
\label{rhocl}
\begin{equation}
\label{rhopluscl}
\rho^+_{\rm cl}(t) = \left( \exp\left[ \frac{t-t_0}{g} \right] + 1 \right)^{-1} \,,
\end{equation}
where again $t_0$ is a collective coordinate. The instanton configuration
joins the two minima of the double-well potential. It does not
define a periodic path whose starting point at $t= -\infty$ is the
same as the end point at $t=\infty$. In order to construct a
periodic orbit, one has to ``join'' two instantons according to
Fig.~5 of Sec.~4.4.1 of Ref.~\cite{ZJJe2004i}, so that the pseudoparticle
returns to its initial point. This can be done by joining the
instanton trajectory $\rho^+_{\rm cl}(t)$ with the trajectory
$\rho^-_{\rm cl}(t)$, where
\begin{equation}
\label{rhominuscl}
\rho^-_{\rm cl}(t) = \left( \exp\left[ \frac{t_0-t}{g} \right] + 1 \right)^{-1} \,,
\end{equation}
\end{subequations}
which is shown in Fig.~\ref{fig1}(d), for the case $t_0 = 0$.
The instanton action is found to be
\begin{equation}
\calS_{\rm dw}[\rho^\pm_{\rm cl}] = \frac{1}{6 \, g} \,.
\end{equation}
Let us recall an important difference of the odd oscillators (and of the double well,
by the way) to the even oscillators. In the former cases, the instanton configuration
is found for all positive, real values of the coupling parameter.
In the latter case (even oscillator), one first has to scale the
action to be consistent with the case of negative coupling
[see Eq.~\eqref{Seven}], before solving for the instanton
[see Eq.~\eqref{chicl} and Fig.~\ref{fig1}(a)].
For the even oscillator, the instanton only exists for
specific regions in the space of complex coupling constants.

%
% Quantum Fluctuations and Decay Widths
%
\section{Quantum Fluctuations and Decay Widths}

According to Refs.~\cite{ZJ1996,ZJ2003},
the quantum fluctuations about the instanton
configurations give rise to a functional determinant
of a Bargmann potential, whose spectrum and determinant
are analytically calculable.
For the ground state of an even oscillator, this calculation
leads to the result
\begin{align}
\label{ImE0}
{\rm Im} \, E_0^{(N)}(g)
\; \mathop{\sim}^{g \to 0^-}\;
-\frac{1}{\sqrt{2 \pi}} \,
\left( \frac{2 \calC(N)}{(-g)^{2/(N-2)}} \right)^{1/2}
\exp\left( -\frac{{\cal A}(N)}{(-g)^{2/(N-2)}} \right) \,,
\end{align}
where
\begin{equation}
\label{Cm}
{\cal C}(m) = \frac{1}{2^{2/(m-2)}} \,
\exp \left[ \; \int\limits_0^{2^{-1/(m-2)}}
\dd q \, \left( \frac{\sqrt{2}}{\sqrt{\WW(m,q)}} - \frac{2}{q} \right) \right]
= \; 2^{2/(m-2)} \,.
\end{equation}
A slight generalization of this result to the $n$th resonance energy
$E_n^{(N)}(g)$, for potentials of even order $N$,
leads to
\begin{equation}
\label{ImEn}
{\rm Im} \, E_n^{(N)}(g)
\; \mathop{\sim}^{g \to 0^-}\;
-\frac{1}{n! \sqrt{2 \pi}} \,
\left( \frac{2 \calC(N)}{(-g)^{2/(N-2)}} \right)^{n + 1/2}
\exp\left( -\frac{ {\cal A}(N) }{(-g)^{2/(N-2)}} \right) \,.
\end{equation}
We take the opportunity to correct a misprint in Eq.~(8) of Ref.~\cite{JeSuZJ2009prl},
where the sign of the mantissa of the exponent
$n+1/2$ was accidentally inverted. This result
is valid in leading order; corrections are of relative order
$g^{2/(N-2)}$. For odd potentials in the normalization given by
Eq.~(\ref{hM}) and for positive coupling, we have for $g \to 0$,
\begin{align}
\label{Imeps0}
{\rm Im} \, \epsilon_0^{(M)}(g)
\; \mathop{\sim}^{g \to 0^+}\;
-\frac{1}{2 \sqrt{2 \pi}} \,
\left( \frac{2 \calC(M)}{g^{1/(M-2)}} \right)^{1/2}
\, \exp\left( -\frac{{\cal A}(M)}{g^{1/(M-2)}} \right) \,.
\end{align}
The generalization to arbitrary $n$ reads
\begin{align}
\label{Imepsn}
& {\rm Im} \, \epsilon_n^{(M)}(g) \; \mathop{\sim}^{g \to 0^+}\;
-\frac{1}{2 n! \sqrt{2 \pi}} \,
\left( \frac{2 \calC(M)}{g^{1/(M-2)}} \right)^{n + 1/2}
\, \exp\left( -\frac{{\cal A}(M)}{g^{1/(M-2)}} \right) \,.
\end{align}
The sign of the imaginary parts of the resonance energies
in Eqs.~\eqref{ImE0},~\eqref{ImEn},~\eqref{Imeps0} and~\eqref{Imepsn}
is negative, as it should be for a resonance energy (as opposed to an
anti-resonance, where the imaginary part of the energy is positive).
As outlined in Refs.~\cite{ZJ1996,ZJ2003}, the sign of the imaginary
part of the energy is somewhat arbitrary as it is determined by
the square root $\sqrt{e_0} = \pm \ii |\sqrt{e_0}|$, where $e_0 < 0$ is the (only)
negative eigenvalue of the matrix that describes the
quantum fluctuations about the instanton configurations.
As we see in the following, the resonance energies
$E_n^{(N)}(g)$ and $\epsilon_n^{(M)}(g)$ actually have cuts
along the negative and the positive real $g$ axis, respectively.
We have to associate either sign of the imaginary part with
a particular side of the branch cut. This becomes clear when we
discuss the dispersion relations fulfilled by the resonance energies.

%
% Dispersion Relations
%
\section{Dispersion Relations}
\label{dispersion}

The dispersion relations formulate connections of the values of the
resonance energies $E_n^{(N)}(g)$ and $\epsilon_n^{(M)}(g)$ of the
even and odd oscillators as a function of $g$ in the
complex $g$ plane. Before we discuss them, let us first
justify the use of the convention $\sqrt{g}\, q^M$ for the
perturbative term in the odd oscillator of degree $M$
[see Eq.~\eqref{hM}].
The dispersion relation for odd oscillators is intimately linked
with the square-root convention for the coupling term.

For illustrative purposes, we consider the Hamiltonian
in the normalization
$h_M(g) = - \half \partial_q^2 + \half q^2 + \sqrt{g}\, q^M$, with odd $M$.
A change in the sign of the
coupling term $\sqrt{g} \to -\sqrt{g}$ can be compensated by a parity
transformation $q \to -q$. So, if $\psi(q)$ is a resonance
eigenfunction of the cubic Hamiltonian, suitably continued into the
complex plane, then $\psi(-q)$ is a resonance eigenfunction
for the cubic Hamiltonian with a sign change of $\sqrt{g}$,
but the same resonance eigenvalue.
We conclude that the resonance spectrum of the cubic
Hamiltonian is invariant under the transformation $\sqrt{g} \to -\sqrt{g}$.

It is thus natural to formulate the cubic Hamiltonian
as a function of $\sqrt{g}$, because once $g$ is specified,
the ambiguity in defining the square root is absorbed into
the invariance of the resonance spectrum of the Hamiltonian
under a sign change of the coupling term. A specification of $g$ alone
suffices to define the spectrum of resonances.
As a function of $g$, the resonance eigenvalues of
$h_M(g)$ have branch cuts along the positive
real axis. For $g = {\rm Re}(g) + {\rm i}\, 0$, the imaginary parts are
negative, whereas for $g = {\rm Re}(g) - {\rm i} \, 0$,
the imaginary parts are
positive. For negative coupling $g = -|g| < 0$,
the resonance eigenvalues of $h_3(g)$,
\begin{equation}
h_3(-|g|) = - \frac12 \, \partial_q^2 + \frac12 \, q^2 \pm \ii |g|^{1/2}\, q^3\,,
\quad g > 0,
\end{equation}
have no imaginary part at all because the cubic Hamiltonian is
${\mathcal PT}$-symmetric for imaginary coupling and has a purely real
spectrum. Therefore the only branch cut of the resonance eigenvalues
of the cubic Hamiltonian is along the positive real axis.
Note that the ${\mathcal PT}$-symmetry of the cubic Hamiltonian here serves
as a property which facilitates our analysis of the structure
of the branch cuts.

In general, the dispersion relation~\cite{BeDu1999}
for the resonance energy eigenvalues $\epsilon^{(M)}_m(g)$
of a general odd oscillator therefore reads
\begin{subequations}
\begin{equation}
\label{dispOdd}
\epsilon_n^{(M)}(g) = n + \frac12 +
\frac{g}{\pi}\, \int\limits_0^{\infty} \dd s \,
\frac{{\rm Im} \, \epsilon_n^{(M)}(s + {\rm i}\, 0)}{s\, (s - g)}\,.
\end{equation}
The subtracted dispersion
relation~\cite{LoMaSiWi1969,BeWu1971,BeWu1973} for the energies $E_n(N,g)$
of the even anharmonic oscillators of degree $N$ is
\begin{equation}
\label{dispEven}
E^{(N)}_n(g) = n + \frac12 -
\frac{g}{\pi} \, \int\limits_{-\infty}^0 {\rm d}s \,
\frac{{\rm Im}\, E^{(N)}_n(s + {\rm i}\, 0)}{s \, (s - g)} \,.
\end{equation}
\end{subequations}
These are subtracted dispersion relations; the term
$n + \half$ fixes the energy level for vanishing coupling.

%
% Generalized Bender--Wu Formulas
%
\section{Generalized Bender--Wu Formulas}
\label{genbenderwu}

The general expressions for the imaginary parts of the
resonance energies given in Eqs.~\eqref{ImEn} and~\eqref{Imepsn},
together with the dispersion relations~\eqref{dispOdd}
and~\eqref{dispEven}, can be used in order to
derive integral representations for the perturbative
coefficients and to derive their leading factorial
growth for large orders of perturbation theory.
Based on Eq.~\eqref{dispEven}, we can expand the integrand
in powers of $g$, and obtain the
coefficients of the perturbation series for the
energy levels of an even oscillator. We find
for the $n$th level of the oscillator of order $N$,
\begin{subequations}
\label{pertEven}
\begin{equation}
E^{(N)}_n(g) \sim \sum_{K=0}^\infty E^{(N)}_{n,K} \, g^K \,,
\qquad
E^{(N)}_{n,K \geq 1} = -\frac{1}{\pi} \, \int\limits_{-\infty}^0 {\rm d}s \,
\frac{{\rm Im}\, E^{(N)}_n(s + {\rm i}\, 0)}{s^{K + 1}} \,.
\end{equation}
Because ${\rm Im}\, E^{(N)}_n(s + {\rm i}\, 0)
< 0$ for $-\infty < s < 0$ and because the integration
variable $s$ is negative along the entire integration contour,
the sign of the perturbative coefficients is
determined according to the formula $(-1)^{K+1}$
for $K \geq 1$.
By contrast, for the $n$th level of an odd anharmonic
oscillator of degree $M$, we have
\begin{equation}
\label{pertOdd}
\epsilon^{(M)}_n(g) \sim  \sum_{K=0}^\infty \epsilon^{(M)}_{n,K} \, g^K \,,
\qquad
\epsilon^{(M)}_{n,K \geq 1} = \frac{1}{\pi} \,
\int\limits_0^{\infty} {\rm d}s \,
\frac{{\rm Im}\, \epsilon^{(M)}_n(s + {\rm i}\,0)}{s^{K + 1}} \,.
\end{equation}
\end{subequations}
We can again argue that ${\rm Im}\, \epsilon^{(M)}_n(s + {\rm i}\,0) < 0$
for $0 < s < \infty$, i.e.~along the entire integration contour.
This implies that the perturbative coefficients
for the odd anharmonic oscillators are all negative,
except for the zeroth-order term (in $g$), which is simply $n + \half$.

Based on Eq.~\eqref{ImEn} and~\eqref{dispEven},
it is a simple matter of integration to calculate
the large-order perturbative expansion
for an arbitrary level of an even oscillator of arbitrary degree,
\begin{align}
\label{leadingEven}
E_{n,K}^{(N)} \sim & \; \frac{ (-1)^{K+1}\, (N-2)}%
{\pi^{3/2} \, n! \, 2^{K + 1 - n}} \;
\Gamma\left( \frac{N - 2}{2} K + n + \half \right)
\left[ B\left(\frac{N}{N-2}, \frac{N}{N-2}\right)
\right]^{-\frac{N-2}{2} K - n - \half} \,,
\quad K \to \infty ,
\end{align}
which is valid up to corrections of relative order $K^{-1}$.
We here confirm the result of Bender and Wu~\cite{BeWu1971}.
Our new result concerns the case of odd oscillators,
where we integrate~\eqref{dispEven} using Eqs.~\eqref{Imepsn}
and find
\begin{align}
\label{leadingOdd}
\epsilon^{(M)}_{n,K} \sim & \; \frac{2 - M}{\pi^{3/2} \, n! \,
2^{2 K + 1 - n}} \; \Gamma\left( (M - 2) K + n + \half \right)
\left[ B\left(\frac{M}{M-2}, \frac{M}{M-2}\right)
\right]^{-(M - 2) K - n - \half}  \,,
\quad K \to \infty .
\end{align}
The formulas for the $K$th order perturbative coefficients
$E_{n,K}^{(N)}$ and $\epsilon^{(M)}_{n,K}$ have been compared to
numerical calculations for a number of example cases, as discussed
below in Sec.~\ref{higher}.

%
% From WKB Expansions to Generalized Quantization Conditions
%
\chapter{From WKB Expansions to Generalized Quantization Conditions}
\label{approach2}

\vspace*{-0.4cm}
\textcolor{light}{ \rule{\textwidth}{0.2cm} }

%
% Basic Formulas
%
\section{Perturbative Expansion}
\label{basic}

The approach outlined in Sec.~\ref{approach1} is
inspired by field-theoretical considerations and
leads to an intuitive understanding of the physics involved
in the derivation of the leading-order factorial
growth of the perturbative coefficients, as given in
Eqs.~\eqref{leadingEven} and~\eqref{leadingOdd}.
However, this formalism does not make use of the additional
simple structure of the problem at hand, namely, the
structure of a one-dimensional nonrelativistic
quantum mechanical oscillator for which semiclassical or
Wentzel--Kramers--Brioullin (WKB) and
perturbative expansions can be derived without any further
technical difficulties. Here, we strive to combine both methods
in order to derive higher-order formulas which grant us
access to phenomenologically relevant, numerically large
correction terms.

To this end, we first investigate the derivation of
higher-order perturbative approximants to the wave functions and
to the energy eigenvalues of the anharmonic oscillators.
Suitable scaling transformations in the original Schr\"{o}dinger
equation corresponding to the Hamiltonians (\ref{HN}) and (\ref{hM})
prove useful in that regard.
For the even oscillator given by Eq.~\eqref{HN}, we start from
\begin{equation}
\left(
-\frac12 \, \frac{\partial^2}{\partial q^2} +
\frac12 \, q^2 + g\, q^N \right) \, \varphi =
E \, \varphi
\end{equation}
and transform according to $q \to g^{-1/(N-2)} \, q$,
which leads to
\begin{equation}
\label{HNscaled}
\left( -\frac{g^{\frac{4}{N-2}}}{2}\, \frac{\partial}{\partial q^2} +
\frac12 \, q^2 + q^N \right) \, \varphi = g^{\frac{2}{N-2}} \, E \, \varphi \,.
\end{equation}
For (\ref{hM}), we start from
\begin{equation}
\left( -\frac12 \, \frac{\partial^2}{\partial q^2} +
\frac12 \, q^2 + \sqrt{g}\, q^M \right) \,
\varphi = E \, \varphi
\end{equation}
and transform according to $q \to g^{-1/(2 (M-2))}\,q$
\begin{equation}
\label{hMscaled}
\left( -\frac{g^{\frac{2}{M-2}}}{2}\, \frac{\partial}{\partial q^2} +
\frac12 \, q^2 + q^M \right) \, \varphi = g^{\frac{1}{M-2}} \, E \, \varphi \,.
\end{equation}
We identify the general coupling parameter
\begin{equation}
\label{defGG}
\GG \equiv \GG(m, g) = g^{X(m)} \,, \qquad
X(m) = {\frac{2 \trunc{m/2} - m + 2}{m-2}} \,,
\end{equation}
which implies that $\GG(M, g) = g^{1/(M-2)}$ for odd $M$ and
$\GG(N, g) = g^{2/(N-2)}$ for even $N$
(here, $\trunc{X}$ is the largest integer smaller than $X$).
With this identification,
Eqs.~(\ref{HNscaled}) and (\ref{hMscaled}) have the general
structure
\begin{align}
\label{Vmq}
& \left( -\frac12 \, \GG^2\, \frac{\partial}{\partial q^2} +
\VV(m,q) \right) \, \varphi = \GG \, E \, \varphi \,,
\nonumber\\[1.377ex]
& \VV(m,q) = \frac12 \, q^2 + q^m \,.
\end{align}
Note that the potential $\VV(m,q)$ differs from the
potential $\WW(m,q)$ used for the calculation of the instanton
configurations, in the sign of the $q^m$ term
[for the definition of $\WW(m,q)$, see Eq.~\eqref{defWm}].

Using the scaled and unified form for the
Schr\"{o}dinger equation given by Eq.~\eqref{Vmq},
the wave functions and
resonance energies are
amenable to a perturbative treatment.
The general formalism has been outlined in Sec.~3 of
Ref.~\cite{ZJJe2004ii} and is recalled here, with a special
emphasis on the anharmonic oscillators.
We first transform the Schr\"{o}dinger equation
to the Riccati equation by setting
\begin{equation}
\label{WKBprep}
\frac{\varphi'(q)}{\varphi(q)} = - \frac{s(q)}{\GG} \,,
\qquad
\frac{\varphi''(q)}{\varphi(q)} =
\frac{s^2(q)}{\GG^2} - \frac{s'(q)}{\GG} \,,
\end{equation}
The Riccati form of the Schr\"{o}dinger equation then reads,
\begin{subequations}
\label{riccati}
\begin{align}
\label{riccati1}
& \GG \, s'(q) - s^2(q) + \UU^2(q) = 0 \,, \\[1.377ex]
\label{riccati2}
& \UU(q) = \sqrt{ 2 \left[ \, \VV(m,q) - \GG \, E \, \right] }\,.
\end{align}
\end{subequations}
The coupling $G$ formally takes the role of $\hbar$ in the
scaled equation.
The function $S(q)$ is proportional to the
logarithmic derivative of the wave function.
It has an implicit dependence on the degree $m$ of the oscillator,
on the coupling $\GG$ and on the energy $E$, which we
suppress in our notation.
We can uniquely decompose $S(q)$ into even-parity and
odd-parity (in $\GG$) components,
\begin{subequations}
\label{evenodd}
\begin{align}
s(q) =& \; s_+(q) + s_-(q) 
= s_+(m, q, E, \GG) + s_-(m, q, E, \GG) \,,
\end{align}
where
\begin{equation}
s_\pm(m, q, - \GG, -E) = \pm s_\pm(m, q, \GG, E) \,.
\end{equation}
\end{subequations}
This leads to the following system of equations,
\begin{subequations}
\label{system}
\begin{align}
\label{systema}
\GG \, s'_-(q) - s_+(q)^2 - s_-(q)^2 +  \UU^2(q) =&\; 0 \,,
\\[1.377ex]
\label{systemb}
\GG \, s'_+(q) - 2 \, s_+(q) \, s_-(q) =&\; 0 \,.
\end{align}
\end{subequations}
From the second of these equations, we have
\begin{equation}
s_-(q) = \frac{\GG}{2} \, \frac{s'_+(q)}{s_+(q)} \,,
\end{equation}
so that
we can write $s$ as a function of $s_+$ only. By virtue of the decomposition
(\ref{evenodd}), this means that
\begin{equation}
\varphi(q) = \frac{1}{\sqrt{ s_+(q) }} \,
\exp\left( - \GG^{-1} \, \int \dd q \, s_+(q) \right) \,.
\end{equation}
If we approximate
$s(q) \approx s_+(q) \approx \sqrt{2 \VV(m,q)}$,
then this is just the textbook version of the
Wentzel--Kramers--Brioullin (WKB) approximation~\cite{LaLi1958vol3}.
The advantage of the above formalism is that it allows
for systematic expansions both for fixed $E$ in powers of $\GG$,
which leads to the perturbative expansion of the energy levels,
but also in powers of $\GG$ at $\GG\,E$ fixed, which is an expansion relevant
a priori for excited states, because $\GG$ is assumed to be small and
$E$ must be large for $\GG\,E$ to attain values of order unity.
The latter approach leads to the WKB expansion.

We start, though, with the perturbative expansion.
Because the wave function for the $n$th excited state has $n$ zeros
on the real line, we can formulate the following
quantization condition,
\begin{equation}
\label{q1}
\frac{1}{2 \pi \ii} \, \oint_C \dd z\, \frac{\varphi'(z)}{\varphi(z)} =
- \frac{1}{2 \pi \ii \, \GG} \, \oint_C \dd z\, s(z) = n \,,
\end{equation}
where $C$ is a contour that encloses all the zeros in the counterclockwise
(mathematically positive)
direction and the definition in Eq.~(\ref{WKBprep}) for the
logarithmic derivative of the wave function has been used.
Equation (\ref{q1}) is the Bohr--Sommerfeld quantization condition.

We now seek a solution of Eq.~(\ref{system}) by expansion in $\GG$
at $E$ fixed, taking advantage of the elimination of $s_-(q)$
via (\ref{systemb}). It is easy to see that the solutions for
$s(q)$ and $s_+(q)$ can be written in terms of the expression
\begin{equation}
\uu(q) = \sqrt{ 2 \, \VV(m,q) }
\end{equation}
alone. Furthermore, the zeros of the denominators of $S(q)$ and
$S_+(q)$ are given exclusively by the zero(s) of $\uu(q)$. Because
of the simple structure of the Riccati equation
\eqref{riccati}, one can write a recursive scheme, within the
perturbative expansion in powers of $g$ at fixed $E$, that leads to
higher-order perturbative approximants which we would like to label
by a lowercase $s$,
\begin{align}
\label{system_s}
s(q) =& \; \sum_{K=0}^\infty \GG^K \, s_K(q) \,, \qquad
s_0(q) = \uu(q) \,, \qquad
s_1(q) = \frac{\uu'(q) - 2 E}{2 \uu(q)} \,,
\nonumber\\[1.377ex]
s_K(q) =& \; \frac{1}{2 \, \uu(q)} \left( s'_{K-1}(q) -
\sum_{l = 1}^{K-1} s_{K-l}(q) \, s_l(q) \right) \,,
\qquad K \geq 2.
\end{align}
To order $\GG^2$, we find for the symmetric part $s_+(q)$ which is
invariant under the exchange $g \to -g$, $E \to -E$,
\begin{align}
\label{splus}
s_+(q) =& \; \uu(q) - G\,\frac{E}{\uu(q)}
+ \GG^2 \, \left( \frac14 \, \frac{\uu''(q)}{\uu^2(q)} -
\frac38 \, \frac{\uu'(q)^2}{\uu^3(q)} - \frac12 \, \frac{E^2}{\uu^3(q)} \right)
+ {\mathcal O}(\GG^3)\,.
\end{align}
This structure implies that the only location relevant for
the singularities in the contour integral (\ref{q1}) is
the origin in the complex $z$-plane.
For even anharmonic oscillators, this is immediately obvious
because the only zero of the function $u(q) = q \, \sqrt{1 + 2 q^N}$
then lies on the real axis ($N$ even).
For odd anharmonic oscillators, the function
$u(q) = q \, \sqrt{1 + 2 q^M}$ has an additional zero on the
real axis at $x_M = -2^{1/M}$. However, the residue at
$x_M$ vanishes.

Because $s_-(q) = \half\, \GG \,
s'_+(q)/s_+(q)$ and $s(q) = s_+(q) + s_-(q)$,
we can finally write the quantization condition (\ref{q1}) as
\begin{equation}
\label{integral1}
- \frac{1}{2 \pi \ii \, \GG} \, \oint_C \dd z\, s_+(z) = n + \frac12 \,.
\end{equation}
Here, $C$ is a contour that encloses only the origin.
The advantage of this formalism is that the function $B(m, E, \GG)$,
defined by
\begin{equation}
\label{BMEG}
B(m, E,\GG) \equiv - \frac{1}{2 \pi \ii \, \GG} \,
\oint_C \dd z\, s_+(z) \,,
\end{equation}
provides for a universal means of determining the perturbative expansion
for an arbitrary excited level, by means of the perturbative
quantization condition
\begin{equation}
\label{pertquant}
B(m, E,\GG) = n + \frac12 \,.
\end{equation}
The rational is to construct higher-order approximants to the
wave function according to Eq.~\eqref{system_s}, then to
find the residue of the higher-order approximants to the
wave function at the origin. The sum of the residues of the higher-order
(in $G$) terms then defines the function $B(m, E,\GG)$.

%
% Example: Perturbative Expansion of the Cubic
%
\section{Example: Perturbative Expansion of the Sextic Potential}
\label{pert_cubic}

The generalized coupling parameter for the sextic
potential with $N=6$ is given by Eq.~\eqref{defGG} as
$G = g^{1/2}$.  The zeroth-order approximant is
\begin{equation}
s_0(q) = u(q) = q\, \sqrt{1 + 2 q^4} \,.
\end{equation}
Entering into the recursive system~\eqref{system_s} and calculating
the residue at $q=0$, we obtain
\begin{align}
\label{B6}
B(6, E, G) = &\; E - G^2 \left( \frac{25}{8} \, E +
\frac{5}{2} \, E^3 \right)
\nonumber\\[0.3ex]
& \; + G^4 \, \left( \frac{21777}{256} \, E
+ \frac{5145}{32} \, E^3
+ \frac{693}{16} \, E^5 \right)
+ {\mathcal O}(G^6)
\nonumber\\[0.3ex]
= &\; E - g \left( \frac{25}{8} \, E +
\frac{5}{2} \, E^3 \right)
\nonumber\\[0.3ex]
& \; + g \, \left( \frac{21777}{256} \, E
+ \frac{5145}{32} \, E^3
+ \frac{693}{16} \, E^5 \right)
+ {\mathcal O}(g^3)
\end{align}
Note, in particular, that the term $-E/u(q)$ in Eq.~\eqref{splus}
leads to the leading term $E + {\mathcal O}(G^2)$ in Eq.~\eqref{B6}
in view of $u(q) = q + {\mathcal O}(q^4)$.
Note also that the term of order $G$ in Eq.~\eqref{splus}
does not produce any contributions to the residue at the origin.
By explicitly solving the equation
$B(6, E_0^{(6)}, G) = \half$ for the ground state
resonance energy $E_0^{(6)}$ order by order in $G^2 = g$,
one obtains the perturbative expansion (as a function of $g$),
\begin{align}
E_0^{(6)} =& \; \frac12 +
\frac{15}{8}\, G^2 -
\frac{3495}{64}\, G^4 +
{\mathcal O}(G^6)
\nonumber\\[0.3ex]
=& \; \frac12 +
\frac{15}{8}\, g -
\frac{3495}{64}\, g^2 +
{\mathcal O}(g^3) \,,
\end{align}
for the ground state of the sextic potential.

%
% From the WKB Expansion to the Characteristic Functions
%
\section{Semiclassical or WKB Expansion}
\label{WKBexpansion}

We now turn our attention to the WKB expansion, which is an expansion in powers
of $G$ for fixed $G\,E$.  {\em A priori}, a natural starting point for the WKB
expansion would be Eq.~\eqref{riccati2}, with $S(q) \approx \UU(q) = \sqrt{ 2
\, \VV(m,q) - 2 \, \GG \, E }$.  However, a second purpose of our construction
of the WKB expansion is the calculation of the instanton action, and of
perturbations thereof. We have already seen in Sec.~\ref{approach1} that the
instanton configuration does not exist for arbitrary values of the coupling
parameter. In the case of an even oscillator, it exists only for negative
coupling $g < 0$ [see Fig.~\ref{fig1}(a)].  In the case of an odd oscillator,
the instanton configuration involves positive values of the coordinates only if
we assume that $\sqrt{g} < 0$ [see Fig.~\ref{fig1}(b)].  For $\sqrt{g} > 0$,
the instanton configuration entails negative values of the coordinate
[see Fig.~\ref{fig1}(c)]. We therefore have to scale the coupling constants
differently as compared to Sec.~\ref{basic},
if we wish to grasp the instanton configuration within our
semiclassical expansion.

Let us start with the case of an even oscillator,
with $g = - |g| < 0$.
Starting from [see Eq.~\eqref{HN}], we obtain
\begin{equation}
\left(
-\frac12 \, \frac{\partial^2}{\partial q^2} +
\frac12 \, q^2 - |g|\, q^N \right) \, \phi =
E \, \phi
\end{equation}
and transform according to $q \to |g|^{-1/(N-2)} \, q$,
which leads to
\begin{equation}
\label{HNscaled2}
\left( -\frac{|g|^{\frac{4}{N-2}}}{2}\, \frac{\partial}{\partial q^2} +
\frac12 \, q^2 - q^N \right) \, \phi = |g|^{\frac{2}{N-2}} \, E \, \phi \,.
\end{equation}
The sign of the $q^N$ term is inverted as compared to
Eq.~\eqref{HNscaled}. For an odd oscillator, we start from
\begin{equation}
\left( -\frac12 \, \frac{\partial^2}{\partial q^2} +
\frac12 \, q^2 + \sqrt{g}\, q^M \right) \,
\phi = E \, \phi
\end{equation}
and transform according to $q \to - g^{-1/[2 (M-2)]}\,q$,
which gives
\begin{equation}
\label{hMscaled2}
\left( -\frac{g^{\frac{2}{M-2}}}{2}\, \frac{\partial}{\partial q^2} +
\frac12 \, q^2 - q^M \right) \, \phi =
g^{\frac{1}{M-2}} \, E \, \phi \,.
\end{equation}
We define the general coupling parameter
\begin{equation}
\label{defTildeGG}
\tGG \equiv \tGG(m, g) = |g|^{X(m)} \,, \qquad
X(m) = {\frac{2 \trunc{m/2} - m + 2}{m-2}} \,,
\end{equation}
where $X(m)$ is given in Eq.~\eqref{defGG}. With this identification,
Eqs.~(\ref{HNscaled2}) and (\ref{hMscaled2}) have the general structure
\begin{align}
\label{Wmq}
& \left( -\frac12 \, \tGG^2\, \frac{\partial}{\partial q^2} +
\WW(m,q) \right) \, \varphi = \tGG \, E \, \varphi \,,
\nonumber\\[1.377ex]
& \WW(m,q) = \frac12 \, q^2 - q^m \,.
\end{align}
Here, we reencounter the potential $\WW(m,q)$ where the
sign of the $q^m$ term is inverted as compared to $\VV(m,q)$.
We set
\begin{equation}
\frac{\phi'}{\phi} = - \frac{\SW(q)}{\tGG} \,,
\end{equation}
where the symbol $\SW$ (as compared to
$S$) denotes the logarithmic derivative of the
scaled wave function for the potential $\WW(m,q)$.
The Riccati form of the Schr\"{o}dinger equation then reads,
in terms of $\SW$,
\begin{subequations}
\label{riccatiW}
\begin{align}
\label{riccatiW1}
& \tGG \, \SW'(q) - \SW^2(q) + \TT^2(q) = 0 \,, \\[1.377ex]
\label{riccatiW2}
& \TT(q) = \sqrt{ 2 \left[ \, \WW(m,q) - \tGG \, E \, \right] }\,.
\end{align}
\end{subequations}
The recursive scheme for calculating higher-order
WKB approximants then reads
\begin{align}
\label{SWrecur}
\SW(q) =& \;  \sum_{K=0}^\infty \tGG^K \, \SW_K(q) \,,
\qquad \SW_0(q) = \TT(q) \,, \qquad
\SW_1(q) = \frac{\SW_0'(q)}{2 \, \SW_0(q)} \,,
\nonumber\\[1.377ex]
\SW_K(q) =& \; \frac{1}{2 \, \SW_0(q)} \left( \SW'_{K-1}(q) -
\sum_{l = 1}^{K-1} \SW_{K-l}(q) \, \SW_l(q) \right) \,.
\end{align}
Unlike the perturbative approximants, which have an
isolated pole at $q=0$, the WKB approximants
have a branch cut, which coincides with the
branch cut of the square root function of the
zeroth-order WKB expansion,
\begin{equation}
\SW_0(q) = \sqrt{ 2 \left[ \, \WW(m,q) - \tGG \, E \, \right] } \,.
\end{equation}
In practical calculations, it turns out to be convenient to define the square
root function so that its branch cut is along the positive real
axis (this aspect will be discussed below). 

Let us here recall the perturbative
quantization condition derived for the perturbative
approximant $s_+(z)$ as given above in
Eqs.~\eqref{integral1} and~\eqref{BMEG},
\begin{equation}
\label{RHS2pii}
\frac{1}{\GG} \, \oint_C \dd z\, s_+(z) =
- 2 \pi \ii \, B(m, E, \GG) =
- 2 \pi \ii \, \left( n + \frac12 \right) \,.
\end{equation}
Here, $C$ was defined as a contour that encircles all poles
of $s_+(z)$ on the real axis in the mathematically
positive sense.

These correspond to the perturbative functions at a particular
value of $\GG$. In order to make the connection to the
formalism given above, we must therefore identify the particular
value of $\tGG$ that corresponds to the same value
of the original coupling parameter $g$ as the value of
$\tGG$ we are using for our calculations.

For even potentials, this identification proceeds as follows.
We have
\begin{equation}
G = g^{2/(N-2)}
\end{equation}
and $\tGG$, for negative $g$, is given by
\begin{equation}
\tGG = |g|^{2/(N-2)} = (-g)^{2/(N-2)} \,, \qquad
g < 0 \,,
\end{equation}
so that
\begin{equation}
\tGG
= \left( - \GG^{(N-2)/2} \right)^{2/(N-2)}
= \exp\left( \frac{2 \pi \ii}{N-2} \right) \, \GG
= \left( -g \right)^{2/(N-2)}  \,,
\end{equation}
on the cut, i.e.~for $g < 0$.
For odd $M$, we have
\begin{equation}
\GG = g^{1/(M-2)}
\end{equation}
and
\begin{equation}
\tGG = |g|^{1/(M-2)}
= g^{1/(M-2)} \,, \qquad
g > 0 \,,
\end{equation}
so that
\begin{equation}
\tGG = \GG = g^{1/(M-2)}\,.
\end{equation}
We thus have
\begin{equation}
\label{Phim}
\tGG = \exp[ \ii \, \Phi(m) ] \; G \,, \qquad
\Phi(m) =
\left\{ \begin{array}{cc}
\displaystyle{\frac{2 \pi}{m-2}} & \qquad \mbox{$m$ even} \\[0.3ex]
1 & \qquad \mbox{$m$ odd}
\end{array} \right. \,.
\end{equation}

Consequently, it appears useful to consider
the contour integral of the WKB expansion around its cut,
\begin{equation}
\label{contour_integral}
\frac{1}{\tGG} \, \oint_{\CC} \dd z\, \SW_+(z)  \,,
\end{equation}
where $\CC$ is a contour that encircles the cut of the WKB
expansion in the mathematically negative, clockwise sense, and
$S_+(z)$ is the part of the WKB approximation $S(z)$
that is symmetric under the transformation $g \to -g$,
$E \to -E$.
Namely, in analogy to~\eqref{evenodd}, we define the decomposition
\begin{subequations}
\label{evenodd2}
\begin{align}
\SW(q) =& \; \SW_+(q) + \SW_-(q) = \SW_+(m, q, E, \GG) + \SW_-(m, q, E, \GG) \,,
\end{align}
where
\begin{equation}
\SW_\pm(m, q, -E, -\GG) = \pm \, \SW_\pm(m, q, E, \GG) \,.
\end{equation}
\end{subequations}
The contour
integral about the cut of the function $\SW_+$  leads to
a more complex structure as compared to
the right-hand side of Eq.~(\ref{integral1}),
\begin{align}
\label{WKB}
& \frac{1}{\tGG} \; \oint\limits_{\CC} \dd z \, \SW_+(q) =
A(m, E, G) + \tfrac12 \ln(2 \pi)
- \ln \left[ \Gamma\left( \half - B(m, E, \GG) \right) \right]
+ B(m, E, \GG) \, \ln\left( - \frac{\tGG}{2 \, {\mathcal C}(m)} \right) \,.
\end{align}
The left-hand
side of the above equation is calculated as a function of $\tGG$,
{\em a priori} [including the WKB approximation to the wave
function, $\SW(q)$]. The right-hand side defines, implicitly,
functions of $\GG$. The procedure is to calculate the integral of
$\SW_+(q)$ about the cut, first of all, as a function of $\tGG$,
\begin{equation}
\label{LHS}
f(\tGG) = \frac{1}{\tGG} \; \oint\limits_{\CC} \dd z \, \SW_+(q) \,,
\end{equation}
and then to identify
\begin{align}
\label{RHS}
& f( \tGG ) =
f\left( \Phi(m) \, \GG \right) =
A(m, E, G) + \tfrac12 \ln(2 \pi)
\nonumber\\[1.377ex]
& \qquad - \ln \left[ \Gamma\left( \half - B(m, E, \GG) \right) \right]
+ B(m, E, \GG) \,
\ln\left( -\frac{\exp[ \ii \, \Phi(m) ] \; \GG}{2 \, {\mathcal C}(m)} \right) \,.
\end{align}
Let us analyze the right-hand side of Eq.~\eqref{RHS}.
To this end, we expand the expression
$\ln\left[ \Gamma\left( \half - B(m, E, \GG) \right) \right]$ for large
principal quantum numbers. Because
\begin{equation}
\label{lnexpanded}
\ln \Gamma(\tfrac12 + z) \approx z\,\ln(z) - z \,,
\end{equation}
and in view of
\begin{equation}
\label{BmEGfirst}
B(m, E,\GG) = E + \calO(G) \,,
\end{equation}
we can approximate the logarithmic terms on the right-hand side
of~\eqref{RHS} as follows.
Taking only the first term from Eq.~\eqref{lnexpanded},
we obtain
\begin{align}
& -\ln \Gamma\left(\half - B(m, E,\GG)\right)
+ B(m, E, \GG) \, \ln\left( - \frac{\tGG}{2 \, {\mathcal C}(m)} \right)
\nonumber\\[1.377ex]
& \qquad \approx B(m, E,\GG) \, \ln(-E)
+ B(m, E, \GG) \, \ln\left( - \frac{\tGG}{2 \, {\mathcal C}(m)} \right)
\approx B(m, E,\GG) \,
\ln\left( \frac{\tGG{} \, E}{2 \, {\mathcal C}(m)} \right) \,.
\end{align}
This expression is real if $\tGG$ is positive (if $E$ is real).
However, if the arguments of both logarithms in the second line of the
above equation acquire an infinitesimal negative imaginary part,
then both logarithms acquire an imaginary part of
$\pm \pi i$, and the term $2 \pi \ii B(m, E, \GG)$,
previously encountered the right-hand side of Eq.~\eqref{RHS2pii},
is recovered, thus establishing a connection of the WKB and the
perturbative contour integrals.

The strategy now is to calculate the contour integral
in Eq.~\eqref{LHS} and then
to use the already determined result
for the function $B(m, E, \GG)$ (see Sec.~\ref{basic} of the
current work) to subtract
perturbative terms from the right-hand side of Eq.~\eqref{RHS},
in order to obtain subsequently
higher-order terms of the function $A(m, E, G)$ in the WKB limit
$\GG \to 0$ and $\GG\, E$ fixed.
A well-defined analytic procedure for the evaluation of the contour integral
in Eq.~\eqref{LHS}, based on the method of asymptotic matching
(see Sec.~7.4 of Ref.~\cite{BeOr1978}),
is presented below in Sec.~\ref{matching}.
In leading order, one can show that
\begin{equation}
\label{leadA}
A(m, E, G) \approx \frac{{\cal A} (m)}{\tGG} =
\left\{ \begin{array}{cc}
\displaystyle{
\frac{{\cal A} (m)}{(-g)^{2/(m-2)}}} & \qquad \mbox{$m$ even} \\[0.3ex]
\displaystyle{
\frac{{\cal A} (m)}{g^{1/(m-2)}}} & \qquad \mbox{$m$ odd}
\end{array} \right. \,,
\end{equation}
where ${\cal A} (m)$ is the instanton action defined in Eq.~\eqref{Am}.
We will therefore call $A(m, E, G)$ the instanton function
in the current article.

At the end of the calculation, it is thus natural to reexpress the
``perturbative $B$ function'' and the ``instanton $A$ function''
in terms of the original coupling $g$ given in Eqs.~(\ref{HN}) and~(\ref{hM}).
We thus define
\begin{subequations}
\label{defABm}
\begin{align}
\label{defAm}
A_m(E,g) =& \; A(m, E, g^{\frac{2 \trunc{m/2} - m + 2}{m-2}} ) \,,
\\[1.377ex]
\label{defBm}
B_m(E,g) =& \; B(m, E, g^{\frac{2 \trunc{m/2} - m + 2}{m-2}} ) \,.
\end{align}
\end{subequations}
In this convention, the $B_m$ functions have an argument of integer
power (just $g$) and contain also only integer powers of the
coupling in their expansion in $g$. The $A_m$ functions,
though, contain fractional powers except for the oscillators of the
third, the fourth, and the sixth degree (the latter one is a special
case).

%
% Example of Quartic Potential
%
\section{Example: Contour Integrals for the Quartic Potential}
\label{matching}

We remember that the contour integral around the cut of the
WKB expansion is to be evaluated for the case where instantons
persist, i.e., for $g < 0$.
We have $N=4$ and therefore, in view of
Eqs.~\eqref{defGG} and~\eqref{defTildeGG},
\begin{equation}
\tGG= -G = - g \qquad \qquad \mbox{$(N=4)$\,.}
\end{equation}
The task is to evaluate the contour integral of the
symmetric part $\SW_+(q)$ of the WKB expansion of the
logarithmic derivative of the wave function,
as defined in Eq.~\eqref{contour_integral}.
We approximate
\begin{align}
\label{defT}
T =& \;
\frac{1}{\tGG} \oint\limits_{\cal C} \dd q\, \SW_+(q)
\approx -\frac{1}{g} \oint\limits_{\cal C} \dd q\, \left[ \SW_0(q)
+ g^2 \, \SW_2(q) + g^4 \, \SW_4(q) \right] = T_0 + T_2 + T_4 \,.
\end{align}
Here, the WKB approximants
$\SW_0(q)$, $\SW_2(q)$ and $\SW_4(q)$ are defined in Eq.~\eqref{SWrecur},
and the integrals $T_0$, $T_2$ and $T_4$ are defined in the
obvious manner.
One can easily show that the contributions from
the terms $\SW_1(q)$ and $\SW_3(q)$ vanish,
and we recall that $\calC$ is a contour that encircles the cut of the WKB expansion
in the clockwise direction. The zeroth-order term is
\begin{equation}
T_0 = -\frac{1}{g} \oint\limits_{\cal C} \dd q \; \SW_0(q) \,,
\end{equation}
with the zeroth-order WKB approximant
\begin{equation}
\SW_0(q) = \sqrt{ 2 \WW(4, q) - \tGG E } = \sqrt{ q^2 - 2 q^4 + g E } \,.
\end{equation}
We define the square root function to have its
branch cut along the positive real axis.
Directly above the cut, the value of the square root is positive,
while directly below, it is negative.
As the contour is clockwise, we thus have an integration interval
from zero to $1/\sqrt{2}$, while below the real axis,
we go from $1/\sqrt{2}$ to zero. With this definition, the
contour integral is
\begin{equation}
T_0 = -\frac{1}{g} \oint\limits_{\cal C} \dd q\,
\sqrt{ q^2 - 2 q^4 + g E }
= -\frac{2}{g} \, \rept
\int\limits_0^{\tfrac{1}{\sqrt{2}}} \dd q\, ( q^2 - 2 q^4 + g E )^{1/2} \,.
\end{equation}
The specification of the real part is necessary because we
evaluate the contour integral about the cut of of the WKB expansion.
For the evaluation, we use the method of asymptotic
matching (see Ch.~7.4 of Ref.~\cite{BeOr1978})
which involves an overlapping parameter $\epsilon$,
that is also used in Lamb shift calculations~\cite{Fe1949,Pa1993,JePa1996} in order
to separate the low-energy from the high-energy contribution
to the bound-electron self-energy.
The overlapping parameter fulfills
\begin{equation}
0 \ll g\,E \ll \epsilon \,,
\end{equation}
and we divide the integration interval $(0, 1/\sqrt{2})$ into three
parts, which are the intervals
$(0, \epsilon)$,
$(\epsilon, \tfrac{1}{\sqrt{2}} - \epsilon)$, and
$(\tfrac{1}{\sqrt{2}} - \epsilon, 1/\sqrt{2})$.
Each result is first expanded in $g\, E$, then in $\epsilon$,
and the divergent terms in $\epsilon$ cancel at the end of the calculation.
We remember that the integral is evaluated for $g < 0$. The integral
$T_0$ is recovered as
\begin{equation}
T_0 = -\frac{2}{g} \, \left( I_1 + I_2 + I_3 \right) \,,
\end{equation}
where the $I_j$ comprise the above mentioned three intervals,
respectively. The first contribution is, up to order $(g\,E)^4$,
\begin{align}
I_1 =& \; \rept \int\limits_0^{\epsilon} \dd q\, ( q^2 - 2 q^4 + g E )^{1/2}
\nonumber\\
=& \;
(g E) \, \left[ \frac12 + \frac12 \, \ln\left( \frac{2 \epsilon^2}{g E} \right) \right]
+ (g E)^2 \,
\left[ \frac78 + \frac{1}{4 \epsilon^2} -
\frac34 \, \ln\left( - \frac{2 \epsilon^2}{g E} \right) \right]
\nonumber\\
& \; + (g E)^3 \, \left[ -\frac{319}{48}
- \frac{1}{8 \epsilon^4} - \frac{5}{4 \epsilon^2} +
\frac{35}{8} \, \ln\left( - \frac{2 \epsilon^2}{g E} \right) \right]
\nonumber\\
& \; + (g E)^4 \,
\left[ \frac{23189}{384}
+ \frac{5}{48 \epsilon^6}
+ \frac{35}{32 \epsilon^4}
+ \frac{315}{32 \epsilon^2}
- \frac{1155}{32} \, \ln\left( - \frac{2 \epsilon^2}{g E} \right) \right] \,.
\end{align}
Here, we have taken into account the fact that only the real part
of the result matters; hence, the argument of the
logarithms reads $-2 \epsilon^2/(g E) > 0$, because $g < 0$.
The second interval leads to
\begin{align}
I_2 =& \; \rept \int\limits_{\epsilon}^{\tfrac{1}{\sqrt{2}}-\epsilon}
\dd q\, ( q^2 - 2 q^4 + g E )^{1/2}
\nonumber\\
=& \; \frac16 +
(g E) \, \left[ \frac12 \, \ln\left( \frac{2}{\epsilon^2} \right) \right]
+ (g E)^2 \,
\left[ \frac54
- \frac{1}{4 \epsilon^2}
- \frac{1}{2^{3/4} \epsilon^{1/2}} -
\frac34 \, \ln\left( \frac{2}{\epsilon^2} \right) \right]
\nonumber\\
& \; + (g E)^3 \, \left[ -\frac{319}{48}
+ \frac{1}{8 \epsilon^4} + \frac{5}{4 \epsilon^2}
+ \frac{1}{6 \, 2^{1/4} \epsilon^{3/2}}
+ \frac{25}{4 \, 2^{3/4} \epsilon^{1/2}}
+ \frac{35}{8} \, \ln\left( \frac{2}{\epsilon^2} \right) \right]
\nonumber\\
& \; + (g E)^4 \,
\left[ \frac{13327}{192}
- \frac{5}{48 \epsilon^6}
- \frac{35}{32 \epsilon^4}
- \frac{315}{32 \epsilon^2}  \right.
\nonumber\\
& \; \qquad \left.
- \frac{1}{8 \, 2^{3/4} \epsilon^{5/2}}
- \frac{175}{96 \, 2^{1/4} \epsilon^{3/2}}
- \frac{6755}{128 \, 2^{3/4} \epsilon^{1/2}}
- \frac{1155}{32} \, \ln\left( \frac{2}{\epsilon^2} \right) \right] \,.
\end{align}
The third interval gives rise to the following results,
\begin{align}
& I_3 = \rept \int\limits_{\tfrac{1}{\sqrt{2}}-\epsilon}^{\tfrac{1}{\sqrt{2}}}
\dd q\, ( q^2 - 2 q^4 + g E )^{1/2}
= (g E)^2 \, \left[ \frac{1}{2^{3/4} \epsilon^{1/2}} \right]
+ (g E)^3 \, \left[
- \frac{1}{6 \, 2^{1/4} \epsilon^{3/2}} \right.
\nonumber\\[0.3ex]
& \quad \left. - \frac{25}{4 \, 2^{3/4} \epsilon^{1/2}} \right]
+ (g E)^4 \,
\left[
\frac{1}{8 \, 2^{3/4} \epsilon^{5/2}}
+ \frac{175}{96 \, 2^{1/4} \epsilon^{3/2}}
+ \frac{6755}{128 \, 2^{3/4} \epsilon^{1/2}}
\right] \,.
\end{align}
Terms that originate from the upper limit
of integration in the third interval are expressed in terms of
half-integer powers of $g\,E$. These terms (an example
is $(g\,E)^{7/2} = (g\,E)^3 \, \sqrt{g\,E}$
involve the square root of negative argument because $g \, E < 0$ and
are thus imaginary (hence, they do not contribute to the
real part of the result).
We have defined the square root to have its branch cut along the
negative real axis, and so the contributions from the
upper and lower branch of the contour $\calC$ cancel for these
terms. Finally, after the cancellation of $\epsilon$, $T_0$ is found as
\begin{align}
& T_0 =
-\frac{1}{3 g} + \left\{ - E + E \, \ln\left( -\frac{g E}{4} \right) \right\}
+ g \, \left[ - \frac{17}{4} E^2-
\frac32 \, E^2 \, \ln\left( -\frac{g E}{4} \right) \right]
\nonumber\\[0.3ex]
&
+ g^2 \, \left[ \frac{59}{2} E^3 +
\frac{35}{4} \, E^3 \, \ln\left( -\frac{g E}{4} \right) \right]
+ g^3 \, \left[ -\frac{49843}{192} E^4 -
\frac{1155}{16} \, E^4 \, \ln\left( -\frac{g E}{4} \right) \right] \,.
\end{align}
For the second term,
\begin{equation}
T_2 = -g \oint\limits_{\cal C} \dd q\, \SW_2(q) \,,
\end{equation}
we find
\begin{align}
& T_2 =
-\frac{1}{24\,E} + g \, \left[ -\frac{35}{24} -
\frac38 \, \ln\left( -\frac{g E}{4} \right) \right]
\nonumber\\[0.3ex]
&
+ g^2 \, \left[ \frac{2093}{96} E +
\frac{85}{16} \, E \, \ln\left( -\frac{g E}{4} \right) \right]
+ g^3 \, \left[ - \frac{16285}{48} E^2 -
\frac{2625}{32} \, E^2 \, \ln\left( -\frac{g E}{4} \right) \right] \,.
\end{align}
The third term is
\begin{equation}
T_4 = -g^3 \oint\limits_{\cal C} \dd q\, \SW_4(q) \,,
\end{equation}
with
\begin{equation}
T_4 = \frac{7}{2880\,E^3} - \frac{3\,g}{640\,E^2} + \frac{821\,g^2}{3840\,E}
+ g^3 \, \left[ -\frac{205621}{4608} -
\frac{1995}{256} \ln\left( -\frac{g E}{4} \right) \right] \,.
\end{equation}
The total result of the contour integral of the WKB expansion is
$T = T_0 + T_2 + T_4$, where
\begin{align}
\label{Tterms}
& T = \frac{7}{2880\,E^3} - \frac{1}{24\,E}
-\frac{1}{3 g} +
\left\{ - E + E \, \ln\left( -\frac{g E}{4} \right) \right\}
\nonumber\\
& + g \,
\left[ - \frac{3}{640\,E^2} - \frac{35}{24} - \frac{17}{4} E^2
+ \left( - \frac38 - \frac32 \, E^2 \right) \, \ln\left( -\frac{g E}{4} \right) \right]
\nonumber\\
&
+ g^2 \, \left[  \frac{821}{3840\,E} + \frac{2093}{96} E + \frac{59}{2} \, E^3
+ \left( \frac{85}{16} E + \frac{35}{4} \, E^3 \right) \, \ln\left( -\frac{g E}{4} \right) \right]
\nonumber\\
& + g^3 \, \left[ -\frac{205621}{4608} - \frac{16285}{48} E^2
- \frac{49843}{192} E^4
+ \left( - \frac{1995}{256} - \frac{2625}{32} \, E^2 - \frac{1155}{16} E^4 \right) \,
\ln\left( -\frac{g E}{4} \right) \right] \,.
\end{align}
The perturbative counterterms read
\begin{align}
\label{Pterms}
& P = \tfrac12 \ln(2 \pi) - \ln \left[ \Gamma\left( \half - B_4(E, g) \right) \right]
+ B_4(E, g) \, \ln\left( \frac{g}{4} \right)
\nonumber\\
& = \frac{7}{2880\,E^3} - \frac{1}{24\,E}
+ \left\{ - E + E \,\ln\left( -\frac{g E}{4} \right) \right\}
\nonumber\\
& + g \,
\left[ - \frac{3\,g}{640\,E^2} - \frac{1}{16}
+ \left( - \frac38 - \frac32 \, E^2 \right) \, \ln\left( -\frac{g E}{4} \right) \right]
\nonumber\\
&
+ g^2 \, \left[  \frac{821}{3840\,E} + \frac{5}{6} E + \frac{9}{8} \, E^3
+ \left( \frac{85}{16} E + \frac{35}{4} \, E^3 \right) \, \ln\left( -\frac{g E}{4} \right) \right]
\nonumber\\
& + g^3 \, \left[ -\frac{4349}{1024} - \frac{1649}{128} E^2 -
\frac{201}{16} E^4 + \left( - \frac{1995}{256} - \frac{2625}{32} \,
E^2 - \frac{1155}{16} E^4 \right) \, \ln\left( -\frac{g E}{4}
\right) \right] \, ,
\end{align}
where the result [for the definition of $B_4(E,g)$, see Eq.~\eqref{defABm}]
\begin{align}
B_4(E, g) =& \; B(4, E, g) = E - g \left( \frac{3}{8} +
\frac{3}{2} \, E^2 \right)
+ g^2 \, \left( \frac{85}{16} \, E + \frac{35}{4} \, E^3 \right)
\nonumber\\[1.377ex]
& \; - \, g^3 \, \left( \frac{1995}{256} +
\frac{2625}{32} \, E^2 + \frac{1155}{16} \, E^4 \right)
\end{align}
has been used, as well as the asymptotic expansion
\begin{equation}
\ln\Gamma( \tfrac12 + z) =
z \, \left\{ \ln(z) -1 \right\} + \tfrac12 \ln(2\pi) -
\frac{1}{24 \, z} + \frac{7}{2880\, z^3} +
\dots \,.
\end{equation}
Note that the logarithmic term $E \, \ln(-g E/4)$ in
the perturbative counterterms [Eq.~\eqref{Pterms}] is obtained
after combining the term $-\ln\Gamma(1/2 - B_4(E,g)) \approx
E \, \ln(-E)$ with the term $B_4(E,g) \ln(g/4) \approx E\,\ln(g/4)$.
The instanton function $A_4(E,g)$ for the quartic potential
is finally found to read
\begin{align}
\label{A4Eg}
& A_4(E,g) = T - P = -\frac{1}{3 \, g} +
g\, \left( -\frac{67}{48} - \frac{17}{4} E^2 \right)  +
g^2\, \left( \frac{671}{32} \, E + \frac{227}{8} E^3 \right)
\nonumber\\[1.377ex]
& \qquad + g^3\, \left( -\frac{372101}{9216}
- \frac{125333}{384} E^2 - \frac{47431}{192} E^4 \right)
+ {\mathcal O}(g^4) \,.
\end{align}
We finally remark that in all intermediate expressions~\eqref{defT}---\eqref{A4Eg},
we have neglected higher-order terms that contribute in the
order $g^4$ and higher to the function $A_4(E,g)$.
The term of order $g^4$ can be found below in Eq.~\eqref{A4}.

Alternative methods for the evaluation of the
contour integrals about the cut of the WKB expansion are
described in Appendix~F of Ref.~\cite{ZJJe2004ii}.
One of these is based on a Mellin transformation with respect
to $E$, after which the $q$ integral can be performed with ease,
followed by a Mellin backtransformation, which evaluates the
original integrals. The second alternative method is an adaptation
of dimensional regularization and is described in
Appendix~F.7 of Ref.~\cite{ZJJe2004ii}. The method described above,
which directly evaluates the integrals along contours
infinitesimally displaced from the real axis, might be most transparent one
among the various approaches, and therefore we have chosen it in the
current context for illustrative purposes.

%
% Generalized Quantization Conditions
%
\section{Generalized Quantization Conditions}
\label{gen_quant}

In Sec.~\ref{basic}, we have investigated the perturbative
expansion for the logarithmic derivative of the
wave function.
The perturbative quantization condition
$B_m(E,g) = n + \tfrac12$ was derived
for an oscillator of degree $m$, where $m$
can be either even or odd
[see Eqs.~\eqref{pertquant} and~\eqref{defABm}].
For the ``stable'' configurations which concern
even potentials with positive coupling and
odd potentials with imaginary coupling ${\rm Re}\,g = 0$,
there are no nontrivial saddle points of the Euclidean
action, and thus no instanton configurations to consider.
The quantization conditions are thus equal to the
perturbative condition, and read
\begin{subequations}
\label{quantStable}
\begin{align}
B_N(E,g) =& \;  n + \tfrac12 \,, \qquad
\mbox{$N$ even} \,, \qquad
g > 0 \,, \\[0.3ex]
B_M(E,g) =& \; n + \tfrac12 \,, \qquad
\mbox{$M$ odd} \,, \qquad
g < 0 \,.
\end{align}
\end{subequations}
The question then is how to modify these conditions
for those values of the coupling constant where
instanton configurations become relevant.
To this end, let us first observe that we can write the
conditions as
\begin{equation}
\frac{1}{\Gamma(\tfrac12 - B_m(E,g))} = 0 \,,
\end{equation}
where $m$ can be either even or odd.
So, on the one hand, according to Eq.~\eqref{BmEGfirst}, we have
$B_m(E,g) = E + \calO(g)$.
On the other hand, the quantity
$1/\Gamma(\tfrac12 - E)$ can naturally be identified as the spectral
determinant of the harmonic oscillator with Hamiltonian
\begin{equation}
{\cal H} = -\frac12 \, \frac{\partial^2}{\partial q^2} + \frac12 \, q^2 \,.
\end{equation}
In this form, we naturally identify
\begin{equation}
\frac{1}{\Gamma(\tfrac12 - E)} = {\rm det}({\cal H}-E)
\end{equation}
as the spectral determinant. The effect of the instanton is to
add an infinitesimal imaginary part to the energy of the bound state,
\begin{equation}
E \approx {\cal E} \equiv n + \tfrac12 + \ii \, \impt \, {\cal E} \, .
\end{equation}
Expanding in the nonperturbatively small (in $g$) imaginary part of the
energy, we obtain
\begin{equation}
\label{seed}
\frac{1}{\Gamma(\tfrac12 - {\cal E})} \approx
%\frac{1}{\Gamma(- n - \ii \, \impt \, {\cal E})} =
- (-1)^{n+1/2} \; n! \;  \impt \, {\cal E}
\end{equation}
Now, starting with an even oscillator (for definiteness),
the imaginary part of the energy can be approximated as
[see Eq.~\eqref{ImEn}]
\begin{align}
\label{gen_im_even}
\impt \, {\cal E} \approx & \;
\impt \, E_n^{(N)}(g) \approx
-\frac{1}{n! \sqrt{2 \pi}} \,
\left( \frac{2 \, \calC(N)}{(-g)^{2/(N-2)}} \right)^{n + 1/2}
\exp\left( -\frac{ {\cal A}(N) }{(-g)^{2/(N-2)}} \right)
\nonumber\\[0.3ex]
\approx & \;
-\frac{1}{n! \sqrt{2 \pi}} \,
\left( \frac{2 \, \calC(N)}{(-g)^{2/(N-2)}} \right)^{B_N(E,g)}
\exp\left(  -A_N(E,g) \right) \,,
\end{align}
we have used~\eqref{leadA}.
If we now generalize the left-hand side of Eq.~\eqref{seed}
as $1/\Gamma(\tfrac12 - B_N(E,g))$ and approximate the
right-hand side of Eq.~\eqref{seed} by the right-hand side
of Eq.~\eqref{gen_im_even}, we have
\begin{align}
\label{quantEven}
& \frac{1}{\Gamma\left( \half - B_N(E,g)\right)} =
\frac{1}{\sqrt{2 \pi}}\,
\left(-\frac{2 \, {\cal C}(N)}{(-g)^{2/(N-2)}} \right)^{B_N(E,g)} \,
\exp\left(  -A_N(E,g) \right) \,, \nonumber\\[0.3ex]
& \mbox{$N$ even} \,, \qquad g < 0 \,,
\end{align}
which is our conjecture for even oscillators in the
``unstable'' regime of negative coupling parameter $g < 0$.

For an odd oscillator, we approximate the right-hand side of
Eq.~\eqref{seed} as
\begin{align}
\label{gen_im_odd}
\impt \, {\cal E} \approx & \;
{\rm Im} \, \epsilon_n^{(M)}(g) \approx
-\frac{1}{n! \sqrt{8 \pi}} \,
\left( \frac{2 \, \calC(M)}{g^{1/(M-2)}} \right)^{n + 1/2}
\, \exp\left( -\frac{{\cal A}(M)}{g^{1/(M-2)}} \right)
\nonumber\\[0.3ex]
\approx & \; -\frac{1}{n! \sqrt{8 \pi}} \,
\left( \frac{2 \, \calC(M)}{g^{1/(M-2)}} \right)^{B_M(E,g)}
\, \exp\left( -A_M(E, g) \right) \,.
\end{align}
We now generalize the left-hand side of Eq.~\eqref{seed}
as $1/\Gamma(\tfrac12 - B_M(E,g))$ and approximate the
right-hand side of Eq.~\eqref{seed} by the right-hand side
of Eq.~\eqref{gen_im_odd}, and obtain
\begin{align}
\label{quantOdd}
& \frac{1}{\Gamma\left( \half - B_M(E, g)\right)} =
\frac{1}{\sqrt{8 \pi}}\,
\left( - \frac{2 \, {\cal C}(M)}{g^{1/(M-2)}} \right)^{B_M(E, g)} \,
\exp[-A_M(E, g)] \,, \nonumber\\[0.3ex]
& \mbox{$M$ odd} \,, \qquad g > 0 \,,
\end{align}
which is our conjecture for odd oscillators in the
``unstable'' regime of positive coupling parameter $g > 0$.
In writing Eq.~\eqref{quantOdd},
we correct a typographical error in Eq.~(18b) of Ref.~\cite{JeSuZJ2009prl},
which concerns the missing minus sign in the
expression $\left( - \frac{2 \, {\cal C}(M)}{g^{1/(M-2)}} \right)^{B_M(E, g)}$.
The quantization conditions~\eqref{quantEven} and~\eqref{quantOdd}
effectively assign specific numerical constants to the expression
on the right-hand side of Eq.~\eqref{WKB}.

%
% Generalized Perturbative Expansions
%
\section{Generalized Perturbative Expansions}

We intend to investigate the structure of generalized perturbative
expansions which fulfill the quantization conditions~\eqref{quantEven}
and~\eqref{quantOdd}. To this end, we enter into the quantization conditions
(\ref{quantEven}) and (\ref{quantOdd}) with an ansatz that
involves an unperturbed energy of the form $n + \half + \eta$ where
$\eta$ is a nonperturbatively small imaginary part that is assumed
to be composed of a nonperturbative exponential factor,
multiplied by fractional powers of the coupling constant.
We then expand the $\Gamma$ functions in
Eqs.~(\ref{quantEven}) and (\ref{quantOdd}) about their
poles and equate the coefficients of the fractional-power terms. In this manner, we easily
derive the following expressions for the decay width
of an arbitrary state of an even or odd anharmonic
oscillator. We thereby recover the results given here in
Eqs.~\eqref{ImEn} and~\eqref{Imepsn}, but we are able to
qualify the order of the relative correction terms. They read
\begin{subequations}
\begin{align}
\label{genimEven}
{\rm Im} \, E_n^{(N)}(g < 0) =& \;
- \frac{1}{n! \sqrt{2 \pi}} \,
\left( \frac{2 {\cal C}(N)}{(-g)^{2/(N-2)}} \right)^{n + \half} \,
\exp\left(- \frac{{\cal A}(N)}{(-g)^{2/(N-2)}}\right)
\nonumber\\[1.377ex]
& \; \times \left[ 1 + {\mathcal O}(g^{2/(N-2)}) \right]
\nonumber\\[1.377ex]
=& \; - \frac{1}{\sqrt{\pi} \, n!} \,
\frac{2^{(n \, N + 1)/(N - 2)}}%
{(-g)^{(2 n + 1)/(N-2)}} \,
\exp\left(- \frac{2^{2/(N-2)} \,
B(\frac{N}{N-2}, \frac{N}{N-2})}{(-g)^{2/(N-2)}}\right)\,
\nonumber\\[1.377ex]
& \; \times
\left[ 1 + {\mathcal O}(g^{2/(N-2)}) \right] \,,
\end{align}
for even oscillators, and for odd oscillators, we obtain
\begin{align}
\label{genimOdd}
{\rm Im} \, \epsilon_n^{(M)}(g > 0) =& \;
- \frac{1}{n! \, \sqrt{8 \pi}} \,
\left( \frac{2 {\cal C}(M)}{g^{1/(M-2)}} \right)^{n + \half} \,
\exp\left(-\frac{{\cal A}(M)}{g^{1/(M-2)}}\right)
\nonumber\\[1.377ex]
& \; \times
\left[ 1 + {\mathcal O}(g^{1/(M-2)}) \right] \,
\nonumber\\[1.377ex]
=& \; - \frac{1}{\sqrt{\pi} \, n!} \,
\frac{2^{[(n - 1)\,M+3]/(M-2)}}%
{g^{(2 n + 1)/(2\,M-4)}} \,
\exp\left(- \frac{2^{2/(M-2)} \,
B(\frac{M}{M-2}, \frac{M}{M-2})}{g^{1/(M-2)}}\right)
\nonumber\\[1.377ex]
& \; \times
\left[ 1 + {\mathcal O}(g^{1/(M-2)}) \right] \,.
\end{align}
\end{subequations}
Using the dispersion relation (\ref{dispEven}), we
recover the result from Eq.~\eqref{leadingEven}
for the leading asymptotics for the perturbative
coefficients of an arbitrary state $n$
of an even oscillator of degree $N$,
\begin{subequations}
\begin{align}
\label{genLOEven}
E_{n, K}^{(N)}
\sim & \; \frac{(-1)^{K+1} \,
(N-2)}{\pi^{3/2} \,  n! \, 2^{K + 1 - n}} \,
\left[ B\left(\frac{N}{N-2},
\frac{N}{N-2}\right) \right]^{- (N - 2) \, K/2 - n - 1/2 }
\nonumber\\[1.377ex]
& \; \times
\Gamma\left( \frac{N-2}{2} \, K + n + \frac12 \right)
\left[ 1 + {\mathcal O}\left( \frac{1}{K}\right) \right] \,,
\end{align}
and clarify that the correction terms are of relative order $1/K$.
The analogous, general result for an arbitrary state $n$ of an
odd oscillator of degree $M$ reads [see Eq.~\eqref{leadingOdd}, we
again clarify the order of the correction terms],

\begin{align}
\label{genLOOdd}
\epsilon_{n, K}^{(M)}
\sim & \; -\frac{M-2}{\pi^{3/2} \, n! \, 2^{2 K + 1 - n}} \,
\left[ B\left(\frac{M}{M-2},
\frac{M}{M-2}\right) \right]^{-(M-2)\, K - n - 1/2}
\nonumber\\[1.377ex]
& \; \times
\Gamma\left[ (M-2)\, K + n + \frac12 \right] \,
\left[ 1 + {\mathcal O}\left( \frac{1}{K}\right) \right]\,.
\end{align}
\end{subequations}

The quantization condition (\ref{quantEven})
implies the following general structure
for the generalized non-analytic expansion (sometimes
called ``resurgent expansion'') characterizing the
resonance energy of the $n$th level of the $N$th anharmonic
oscillator (even $N$),
\begin{align}
\label{genEvenHigher}
& E_n^{(N)}(g < 0) =
\sum_{J=0}^\infty
\left[
\frac{\ii}{n! \sqrt{2 \pi}} \,
\left( \frac{2 {\cal C}(N)}{(-g)^{2/(N-2)}} \right)^{n + \half} \,
\exp\left(- \frac{{\cal A}(N)}{(-g)^{2/(N-2)}}\right)
\right]^J
\nonumber\\[1.377ex]
& \; \times \sum_{L=0}^{L_\mathrm{max}}
\, \ln^L\left( -\frac{ 2 {\mathcal C}(N) }{ (-g)^{2/(N-2)}}  \right)
\sum_{K=0}^\infty
\Xi^{(N,n)}_{J,L,K}
\, (-g)^{2 K/(N-2)}
\nonumber\\[1.377ex]
=& \; \sum_{J=0}^\infty
\left[ \frac{\ii}{\sqrt{\pi} \, n!} \,
\frac{2^{(n \, N + 1)/(N - 2)}}%
{(-g)^{(2 n + 1)/(N-2)}} \,
\exp\left(- \frac{2^{2/(N-2)} \,
B(\frac{N}{N-2}, \frac{N}{N-2})}{(-g)^{2/(N-2)}}\right)
\right]^J
\nonumber\\[1.377ex]
& \; \times \sum_{L=0}^{L_\mathrm{max}}
\, \ln^L\left( -\frac{ 2^{N/(N-2)} }{ (-g)^{2/(N-2)}}  \right)
\sum_{K=0}^\infty
\Xi^{(N,n)}_{J,L,K}
\, (-g)^{2 K/(N-2)} \,,
\end{align}
with constant coefficients $\Xi^{(N,n)}_{JLK}$
and $L_\mathrm{max} \equiv \mathrm{max}(0, J-1)$.
The above triple expansion is complicated and in need of an interpretation.
The term with $J = 0$ recovers the basic, perturbative expansion
which has only integer powers in $g$. Therefore,
$\Xi^{(N,n)}_{0,0,K(N-2)/2} = (-1)^K \, E^{(N)}_{n,K}$
where the $E^{(N)}_{n,K}$ represent the perturbative coefficients
from Eq.~(\ref{pertEven}).
Note that the perturbative contribution to the energy levels
is present for both positive and negative coupling $g$.
The term with $J = 1$ recovers the leading contribution
in the expansion in powers of $\exp(-{\mathcal A}(N)/g)$
to the imaginary part of the resonance energy.
The term with $J = 2$ involves a logarithm of the
form $\ln\left( -\frac{ 2 {\mathcal C}(N) }{ (-g)^{2/(N-2)}} \right)$.
The explicit imaginary parts of the logarithms cancel
against the imaginary parts of lower-order perturbation series
which are summed in complex directions, in a manner consistent
with the analytic continuation of the
logarithm~\cite{FrGrSi1985,Je2001pra,JeZJ2001,ZJJe2004ii}.

Let us illustrate the expansion (\ref{genEvenHigher})
by calculating the first terms in the expansion in powers
of $J$ up to $J=3$, including the logarithmic terms, but only up to
leading order in $g$, i.e.~only the terms with $K=0$,
for the ground state of the
quartic oscillator. The result of this calculation is
\begin{align}
& E_0^{(4)}(g < 0) =
\left\{ \frac12 + {\mathcal O}(g) \right\}
- \ii \, \sqrt{ -\frac{2}{\pi \, g} } \,
\exp\left( \frac{1}{3\,g} \right)  \,
\left[ 1 + {\mathcal O}(g) \right]
\nonumber\\[1.377ex]
& +
\frac{ 2 }{\pi \, g} \,
\exp\left( \frac{2}{3\,g} \right)  \,
\left[ \ln\left(\frac{4}{g}\right) + \gamma_{\mathrm E} \right]
\left\{ 1 + {\mathcal O}(g \, \ln(g)) \right\}
\nonumber\\[1.377ex]
& + \ii \, \left( -\frac{2}{\pi \, g} \right)^{3/2} \,
\exp\left( \frac{1}{g} \right)  \,
\left[ \frac32 \,
\left( \ln\left(\frac{4}{g}\right) + \gamma_{\mathrm E} \right)^2 +
\frac12 \,\zeta(2) \right] \,
\left\{ 1 + {\mathcal O}(g \, \ln(g)) \right\} \,.
\end{align}
Here, $\gamma_{\mathrm E}$ is Euler's constant, and
$\zeta(2) = \pi^2/12$ is the zeta function.
The terms with $J = 1$ and $J = 3$ contribute to the imaginary
part of the resonance energy, whereas the terms with
$J = 2$ adds a nonperturbative correction to the real part
of the resonance energy which is present only for $g < 0$.
For small coupling, the imaginary part is of course dominated by the
leading term with $J = 1$, and the corrections, expressible in
powers of $g$, to the $(J = 1)$-term are the phenomenologically
most important ones. These correspond to the coefficients
labelled $\Xi^{(4,0)}_{1,0,K}$ and are given in Eq.~(\ref{im4}) below.

We also note that the
quantization condition (\ref{quantOdd})
implies the following structure
for the generalized non-analytic expansion
characterizing the
resonance energy of the $n$th level of the $M$th odd anharmonic
oscillator,
\begin{align}
\label{genOddHigher}
& \epsilon_0^{(M)}(g > 0) = \;
\sum_{J=0}^\infty
\left[
\frac{\ii}{n! \, \sqrt{8 \pi}} \,
\left( \frac{2 {\cal C}(M)}{g^{1/(M-2)}} \right)^{n + \half} \,
\exp\left(-\frac{{\cal A}(M)}{g^{1/(M-2)}}\right)
\right]^J
\nonumber\\[1.377ex]
& \; \quad \times \sum_{L=0}^{L_\mathrm{max}}
\, \ln^L\left( -\frac{ 2 {\mathcal C}(M) }{ g^{1/(M-2)}}  \right)
\sum_{K=0}^\infty
\Xi^{(M,n)}_{J,L,K} \, g^{K/(M-2)} \,
\nonumber\\[1.377ex]
& = \; \sum_{J=0}^\infty
\left[ \frac{\ii}{\sqrt{\pi} \, n!} \,
\frac{2^{[(n - 1)\,M+3]/(M-2)}}%
{g^{(2 n + 1)/(2\,M-4)}} \,
\exp\left(- \frac{2^{2/(M-2)} \,
B(\frac{M}{M-2}, \frac{M}{M-2})}{g^{1/(M-2)}}\right)
\right]^J
\nonumber\\[1.377ex]
& \; \quad \times \sum_{L=0}^{L_\mathrm{max}}
\, \ln^L\left( -\frac{ 2^{M/(M-2)} }{ g^{1/(M-2)}}  \right)
\sum_{K=0}^\infty
\Xi^{(M,n)}_{J,L,K} \, g^{K/(M-2)} \,.
\end{align}
Here, the perturbation series is recovered as
$\Xi^{(M,n)}_{0,0,K(M-2)} = \epsilon^{(M)}_{n,K}$,
and again $L_\mathrm{max} = \mathrm{max}(0, J-1)$.
Just as for the even oscillators,
it is instructive to illustrate the structure of
this expansion by calculating the first few terms,
restricted to the zeroth order in $g$, i.e.~only the terms
with $K=0$, for the ground state of the cubic oscillator.
The result is
\begin{align}
& \epsilon_0^{(3)}(g > 0) =
\left\{ \frac12 + {\mathcal O}(g) \right\}
- \frac{\ii}{\sqrt{ \pi g}} \,
\exp\left( - \frac{2}{15\,g} \right)  \,
\left[ 1 + {\mathcal O}(g) \right]
\nonumber\\[1.377ex]
& \; - \frac{ 1 }{\pi g} \,
\exp\left( - \frac{4}{15\,g} \right)  \,
\left[ \ln\left(-\frac{8}{g}\right) + \gamma_{\mathrm E} \right]
\left\{ 1 + {\mathcal O}(g \ln(g)) \right\}
\nonumber\\[1.377ex]
& \; + \frac{\ii }{ \left(\pi g\right)^{3/2}} \,
\exp\left( -\frac{2}{5\,g} \right)  \,
\left[ \frac32 \,
\left( \ln\left(- \frac{8}{g}\right) + \gamma_{\mathrm E} \right)^2 +
\frac12 \,\zeta(2) \right] \,
\left\{ 1 + {\mathcal O}(g \, \ln(g)) \right\} \,.
\end{align}
For the perturbative expansion about the first instanton,
see Eq.~(\ref{im3}) below.

%
% Higher--Order Results
%
\chapter{Higher--Order Results}
\label{higher}

\vspace*{-0.4cm}
\textcolor{light}{ \rule{\textwidth}{0.2cm} }

%
% Orientation
%
\section{Orientation}

In the previous chapters of this article, we have derived the
leading-order results for the decay widths and the factorial
asymptotics of the perturbative coefficients for even and
odd anharmonic oscillators. Our formulas are applicable to
general resonances of anharmonic oscillators of arbitrary
degree. In the current chapter, we would like to apply
the generalized quantization conditions (\ref{quantEven})
and~(\ref{quantOdd})
in order to derive concrete results for the higher-order
corrections to the leading-order results for the
first six anharmonic oscillators: those of degrees
$3, 4, \dots, 8$ (cubic, quartic, quintic, sextic, septic and octic
oscillators). The material presented in the
following sections may seem nothing more than a
collection of formulas; yet, the evaluation of the higher-order terms
is somewhat nontrivial because one has to enter the quantization
conditions with an appropriate ansatz for the energy in
order to derive the higher-order terms. We thus felt that it would
be useful to indicate higher-order terms for various
potentials in order to illustrative the applicability of the
new results.

%
% Results for the Cubic Potential
%
\section{Cubic Potential}
\label{cubic}

We recall that we use the cubic
Hamiltonian in the normalization:
$h_3(g) \, = \, -\frac{1}{2} \,
\frac{\partial^2}{\partial q^2}  + \frac{1}{2} \, q^2 +
\sqrt{g} \, q^3$, according to Eq.~(\ref{hM}).
The result for $B_3(E,g)$ is easily derived,
\begin{align}
\label{B_series_cubic}
& B_3(E, g) = E + g \left( \frac{7}{16} +
\frac{15}{4} \, E^2 \right)
+ g^2 \, \left( \frac{1365}{64} \, E + \frac{1155}{16} \, E^3 \right)
\\[1.377ex]
& + \, g^3 \, \left( \frac{119119}{2048} +
\frac{285285}{256} \, E^2 + \frac{255255}{128} \, E^4 \right)
 + g^4 \, \left( \frac{156165009}{16384} \,E \right.
\nonumber\\[1.377ex]
& \left. \quad +
\frac{121246125}{2048} \, E^3 +
\frac{66927861}{1024} \, E^5 \right)
+ \, g^5 \, \left( \frac{10775385621}{262144} +
\frac{67931778915}{65536} E^2 \right.
\nonumber\\[1.377ex]
& \left. +  \frac{51869092275}{16384} \, E^4 +
\frac{9704539845}{4096} \, E^6 \right) + \mathcal{O}(g^6) \, .
\nonumber
\end{align}
The nonalternating perturbation series for the ground state is obtained
as a solution of $B_3(\epsilon_0^{(3)}(g), g) = \half$:
\begin{eqnarray}
\label{P_series_c}
\epsilon_0^{(3)}(g) &=& \frac12 -
\frac{11}{8}\, g -
\frac{465}{32}\, g^2 -
\frac{39709}{128}\, g^3 -
\frac{19250805}{2048}\, g^4
+ \mathcal{O}(g^5) \, .
\end{eqnarray}
The higher-order terms of the $A$ function read
\begin{align}
& A_3(E,g) = \frac{2}{15\,g} +
g\, \left( \frac{77}{32} + \frac{141}{8} \, E^2 \right) +
g^2\, \left( \frac{15911}{128} \, E + \frac{11947}{32} \, E^3 \right)
\\[1.377ex]
& \; + g^3\, \left( \frac{49415863}{122880} +
\frac{6724683}{1024}\, E^2 +
\frac{5481929}{512}\, E^4 \right)
+ g^4\, \left( \frac{2072342055}{32768} \, E \right.
\nonumber\\[1.377ex]
& \; \quad \left. +
\frac{44826677}{128}\, E^3 +
\frac{733569789}{2048}\, E^5 \right)
+ g^5\, \left( \frac{404096853629}{1310720}  \right.
\nonumber\\[1.377ex]
& \; \left. + \frac{1100811938289}{163840}\, E^2 +
\frac{307346388279}{16384}\, E^4 +
\frac{134713909947}{10240}\, E^6 \right)
+ {\mathcal O}(g^6) \,.
\nonumber
\end{align}
With ${\cal C}(3) = 4$,
the conjectured quantization condition for $g > 0$, see
Eq.~(\ref{quantOdd}), reads for the cubic potential,
\begin{equation}
\frac{1}{\Gamma\left( \half - B_3(E,g)\right)} =
\frac{1}{\sqrt{8 \pi}}\, \left( - \frac{8}{g} \right)^{B_3(E,g)} \,
\exp(-A_3(E,g)) \,.
\end{equation}
The nonperturbative terms can now be determined on the basis
of the quantization condition. The solution for the ground state,
which generalizes formulas found by Alvarez~\cite{Al1988},
reads,
\begin{align}
\label{im3}
& {\rm Im} \, \epsilon_0^{(3)}(g > 0) =
-\frac{\exp\left( - \frac{2}{15\,g} \right)}%
{\sqrt{\pi \, g}} \,
\left\{ 1 - \frac{169}{16} \, g -
\frac{ 44507 }{ 512 } \, g^2 -
\frac{ 86071851 }{ 40960 } \, g^3   \right.
\nonumber\\[1.377ex]
& - \frac{ 189244716209 }{ 2621440 } \, g^4
- \frac{ 128830328039451 }{ 41943040 } \, g^5  -
\frac{ 1027625748709963623 }{ 6710886400 } \, g^6
\nonumber\\[1.377ex]
& - \frac{ 933142404651555165943 }{ 107374182400 } \, g^7  -
\frac{ 7583898146256325425743381 }{ 13743895347200 } \, g^8
\nonumber\\[1.377ex]
& - \frac{ 42597573240436340390869984311 }%
{ 1099511627776000 } \, g^9
\nonumber\\[1.377ex]
&  - \frac{ 104925715318980163875631509722177 }%
{ 35184372088832000 } \, g^{10}
\nonumber\\[1.377ex]
& \left. - \frac{ 140719287091496360714368165944515013 }%
{ 562949953421312000 } \, g^{11} + {\mathcal O}(g^{12}) \right\} \,.
\end{align}
In terms of diagrammatic perturbation theory about the
instanton configuration, this result would correspond
to a 22-loop evaluation of the fluctuations about the
instanton (see also Sec.~\ref{numdiag} below).
This translates into the following corrections to the leading
factorial growth of the coefficients,
by virtue of Eq.~(\ref{dispOdd}),
\begin{align}
\label{LO3}
\epsilon_{0,K}^{(3)} =& -\frac{1}{\pi^{3/2}} \,
\left( \frac{15}{2} \right)^{K + \half} \,
\Gamma\left( K + \half \right) \,
\left\{ 1
- \frac{169}{120\, K}
- \frac{ 64787 }{ 28800 \, K^2 }
- \frac{ 48521417 }{ 5760000 \, K^3 } \right.
\nonumber\\[1.377ex]
& - \frac{ 418278626369}{ 8294400000 \, K^4 }
- \frac{ 130056961585097 }{ 331776000000 \, K^5 }
- \frac{ 1459225926418669541 }{ 398131200000000 \, K^6 }
\nonumber\\[1.377ex]
& - \frac{ 5687653597839945011623 }{ 143327232000000000 \, K^7 }
- \frac{ 66875677760063253018761141 }{ 137594142720000000000 \, K^8 }
\nonumber\\[1.377ex]
& - \frac{ 60795686738650254719443177279 }%
{ 9172942848000000000000 \, K^9 }
\nonumber\\[1.377ex]
& - \frac{ 1906445431654625736765273984543497 }%
{ 19813556551680000000000000 \, K^{10} } +
\nonumber\\[1.377ex]
& \left. - \frac{ 22298642113252679371826228458403 }%
{ 15850845241344000000000 \, K^{11} } +
{\mathcal O}\left( \frac{1}{K^{12}} \right) \right\} \,.
\end{align}
This result confirms, and extends by five orders in the $1/K$ expansion,
the result given in Eq.~(11) of Ref.~\cite{BeDu1999}.
For the first excited state, we have for the
decay width up to relative order $g^8$,
\begin{align}
\label{im3level1}
& {\rm Im} \, \epsilon_1^{(3)}(g) =
- \frac{ 8 \, {\rm e}^{-2/(15\,g)} }{ \sqrt{\pi} \, g^{3/2}} \,
\left\{
1 - \frac{853}{16} \, g
+ \frac{33349}{512} \, g^2 - \frac{395368511}{40960} \, g^3
\right.
\nonumber\\[1.377ex]
& \; - \frac{ 1788829864593 }{ 2621440 } \, g^4
- \frac{ 2121533029723423 }{ 41943040 } \, g^5
- \frac{ 27231734458812207783 }{ 6710886400 } \, g^6 +
\nonumber\\[1.377ex]
& - \frac{ 37583589061337851179291 }{ 107374182400 } \, g^7
- \frac{ 442771791224240926548268373 }{ 13743895347200 } \, g^8
\nonumber\\[1.377ex]
& \;
- \frac{ 3458498643307039634422016023691 }{ 1099511627776000 } \, g^9
\nonumber\\[1.377ex]
& \; - \frac{ 11415793836918299403741902225295009 }%
{ 35184372088832000 } \, g^{10}
\nonumber\\[1.377ex]
& \;
\left.
- \frac{ 19846244647223333437533744190161969649 }{ 562949953421312000 }\,
g^{11} + {\mathcal O}(g^{11}) \right\} \,,
\end{align}
where we observe the numerically significant contribution from the
next-to-leading order correction
term $-\frac{853}{16} \, g$, even at small $g$.
The first few correction terms have recently been given
in~\cite{JeSuZJ2009prl}.
The asymptotics
for the perturbative coefficients (large $K$) are
\begin{align}
\label{LO3level1}
\epsilon^{(3)}_{1,K} =& \;
- \frac{ 15^{ K + 3/2 } }{ 2^{ K - 3/2 } \pi^{3/2} } \,
\Gamma\left(K + \tfrac{3}{2} \right) \,
\left\{ 1
- \frac{ 853 }{ 120 \, K }
+ \frac{ 135709 }{ 28800 \, K^2 }
- \frac{ 426076511 }{ 17280000 \, K^3 }
\right.
\nonumber\\[1.377ex]
& \; - \frac{ 687908048171 }{ 2764800000 \, K^4 }
- \frac{ 3037546625512063 }{ 995328000000 \, K^5 }
- \frac{ 16549184414304538261 }{ 398131200000000 \, K^6 }
\nonumber\\[1.377ex]
& \; - \frac{ 29911997493767498508937 }%
{ 47775744000000000 \, K^7 }
- \frac{ 1417974891656576662560155573 }%
{ 137594142720000000000 \, K^8 }
\nonumber\\[1.377ex]
& \; - \frac{ 15158487931720896294670892498891 }%
{ 82556485632000000000000 \, K^9 }
\nonumber\\[1.377ex]
& \; - \frac{ 23212260590744971348988221615146323 }%
{ 6604518850560000000000000 \, K^{10} }
\nonumber\\[1.377ex]
& \; \left. - \frac{ 170817014684450099129620368822347752849 }%
{ 2377626786201600000000000000 \, K^{11} }
+ {\mathcal O}\left( \frac{1}{K^{12}} \right) \right\} \,.
\end{align}
%

%
% Results for the Quartic Potential
%
\section{Quartic Potential}
\label{quartic}

For the quartic
Hamiltonian in the normalization
$H_4(g) = -\frac{1}{2} \,
\frac{\partial^2}{\partial q^2}  + \frac{1}{2} \, q^2 +  g \, q^4$,
according to Eq.~(\ref{HN}),
the result for $B_4(E,g)$ reads
\begin{align}
\label{B_series_q}
B_4(E, g) =& \; E - g \left( \frac{3}{8} +
\frac{3}{2} \, E^2 \right)
+ g^2 \, \left( \frac{85}{16} \, E + \frac{35}{4} \, E^3 \right)
\\[1.377ex]
& \; - \, g^3 \, \left( \frac{1995}{256} +
\frac{2625}{32} \, E^2 + \frac{1155}{16} \, E^4 \right)
\nonumber\\[1.377ex]
& \; + g^4 \, \left( \frac{400785}{1024} \,E +
\frac{165165}{128} \, E^3 +
\frac{45045}{64} \, E^5 \right)
+ \mathcal{O}(g^5) \, .
\nonumber
\end{align}
The alternating perturbation series for the ground state is a solution of
$B_4(E_0^{(4)}(g), g) = \half$,
\begin{equation}
\label{P_series_q}
E_0^{(4)}(g) = \frac12 +
\frac{3}{4}\, g -
\frac{21}{8}\, g^2 +
\frac{333}{16}\, g^3 -
\frac{30885}{128}\, g^4 +
\frac{916731}{256}\, g^5
+ \mathcal{O}(g^6) \, .
\end{equation}
The new result concerns higher-order terms of the $A$ function,
\begin{align}
\label{A4}
A_4(E,g) =& \; -\frac{1}{3\,g} -
g\, \left( \frac{67}{48} + \frac{17}{4} \, E^2 \right) +
g^2\, \left( \frac{671}{32} \, E + \frac{227}{8} \, E^3 \right)
\\[1.377ex]
& \; - g^3\, \left( \frac{372101}{9216} +
\frac{125333}{384}\, E^2 +
\frac{47431}{192}\, E^4 \right)
\nonumber\\[1.377ex]
& \; + g^4\, \left( \frac{3839943}{2048} \, E +
\frac{82315}{16}\, E^3 +
\frac{317629}{128}\, E^5 \right)
+ {\mathcal O}(g^5) \,.
\nonumber
\end{align}
The conjectured quantization condition for an even potential
at negative coupling $g < 0$, see Eq.~(\ref{quantEven}),
becomes
\begin{equation}
\frac{1}{\Gamma\left( \half - B_4(E,g)\right)} =
\frac{1}{\sqrt{2 \pi}}\, \left( \frac{4}{g} \right)^{B_4(E,g)} \,
\exp(-A_4(E,g)) \,.
\end{equation}
The higher-order corrections to the decay width of the ground-state
resonance energy for $g < 0$ can now easily be determined, and we
again present a 22-loop result (in language of Feynman diagrams, see
Sec.~\ref{numdiag}),
\begin{align}
\label{im4}
& {\rm Im} \, E_0^{(4)}(g < 0) = - \exp\left( \frac{1}{3\,g} \right)\,
\sqrt{- \frac{2}{\pi g} } \,
\left( 1
+ \frac{95}{24} \, g
- \frac{ 13259 }{ 1152 } \, g^2
+ \frac{ 8956043 }{ 82944 } \, g^3 \right.
\nonumber\\[1.377ex]
& \; - \frac{ 11481557783 }{ 7962624 } \, g^4
+ \frac{ 4580883830443 }{ 191102976 } \, g^5
- \frac{ 12914334973382407 }{ 27518828544 } \, g^6
\nonumber\\[1.377ex]
& \; + \frac{ 6938216714164463905 }{ 660451885056 } \, g^7
- \frac{ 33483882026182043052421 }{ 126806761930752 } \, g^8
\nonumber\\[1.377ex]
& \; + \frac{ 201610633027633355940784741 }%
{ 27390260577042432 } \, g^9
\nonumber\\[1.377ex]
& \; - \frac{ 296425081490007428403791901173 }{ 1314732507698036736 } \, g^{10}
\nonumber\\[1.377ex]
& \; \left. + \frac{ 237708867821086878016755601858687 }%
{ 31553580184752881664 } \, g^{11}
+ {\mathcal O}(g^{12}) \right) \,.
\end{align}
The corrections to the leading
factorial asymptotics of the coefficients thus read,
in view of Eq.~(\ref{dispEven}),
up to tenth order,
\begin{align}
\label{LO4}
& E_{0,K}^{(4)} = (-1)^{K+1} \, \sqrt{\frac{6}{\pi^3}} \,
  3^K \, \Gamma\left(K+ \tfrac{1}{2}\right) \,
\left\{ 1 - \frac{95}{72\, K} -
\frac{ 20099 }{ 10368 \, K^2 } -
\frac{ 15422651 }{ 2239488 \, K^3 } \right.
\nonumber\\[1.377ex]
& - \frac{ 25875616775 }{ 644972544 \, K^4 } -
\frac{ 14187855712747 }{ 46438023168 \, K^5 } -
\frac{ 56333049074528335 }{ 20061226008576 \, K^6 } -
\nonumber\\[1.377ex]
& - \frac{ 43291321970558012401 }{ 1444408272617472 \, K^7 } -
\frac{ 301836286153388103809317 }{ 831979165027663872 \, K^8 } -
\nonumber\\[1.377ex]
& - \frac{ 2641578170583411485710072165 }%
{ 539122498937926189056 \, K^9 }
- \frac{ 5664009914699943317155936358237 }%
{ 77633639847061371224064 \, K^{10} }
\nonumber\\[1.377ex]
& \left. - \frac{ 6633546006727980489585005984230639 }%
{ 5589622068988418728132608 \, K^{11} }
+ {\mathcal O}\left( \frac{1}{K^{12}} \right) \right\} \,.
\end{align}
For the first excited state, the decay width incurred for
negative coupling reads, up to relative order $g^3$,
\begin{align}
\label{im4level1}
& {\rm Im} \, E_1^{(4)}(g < 0) = - \exp\left( \frac{1}{3\,g} \right)\,
\sqrt{- \frac{32}{\pi g^3} } \,
\left( 1
+ \frac{371}{24} \, g
- \frac{ 3371 }{ 1152 } \, g^2
+ \frac{ 33467903 }{ 82944 } \, g^3
\right.
\nonumber\\[1.377ex]
& - \frac{ 73699079735 }{ 7962624 }  \, g^4
+ \frac{ 44874270156367 }{ 191102976 }  \, g^5
- \frac{ 181465701024056263 }{ 27518828544 }  \, g^6
\nonumber\\[1.377ex]
& + \frac{ 133606590325852428349 }{ 660451885056 }  \, g^7
- \frac{ 850916613482026035123397 }{ 126806761930752 }  \, g^8
\nonumber\\[1.377ex]
& + \frac{ 6544779430184823240660057985 }%
{ 27390260577042432 } \, g^9
\nonumber\\[1.377ex]
& - \frac{ 11946916861319310964857723252821 }%
{ 1314732507698036736 }  \, g^{10}
\nonumber\\[1.377ex]
& \left. + \frac{ 11601477714905759444960434746130339 }%
{ 31553580184752881664 } \, g^{11}
+ {\mathcal O}(g^{12}) \right) \,.
\end{align}
We can now use Eq.~(\ref{dispEven}),
to infer the corrections to the leading
factorial asymptotics of the perturbative coefficients,
\begin{align}
\label{LO4level1}
& E_{1,K}^{(4)} = (-1)^{K+1} \, \sqrt{\frac{32}{\pi^3}} \,
  3^{K + 3/2} \, \Gamma\left(K+ \tfrac{3}{2} \right) \,
\left\{ 1 - \frac{371}{72\, K} -
\frac{ 23341 }{ 10368 \, K^2 }
\right.
\nonumber\\[1.377ex]
& - \frac{ 36352799 }{ 2239488 \, K^3 } - \frac{ 87794214599 }{
644972544 \, K^4 } - \frac{ 67849536512431 }{ 46438023168 \, K^5 } -
\frac{ 356633980356930271 }{ 20061226008576 \, K^{6} }
\nonumber\\[1.377ex]
& - \frac{ 350378589485035426237 }{ 1444408272617472 \, K^7 }
- \frac{ 3041825321871313559291173 }{ 831979165027663872 \, K^8 }
\nonumber\\[1.377ex]
& - \frac{ 32445494418969664996929544513 }%
{ 539122498937926189056 \, K^9 }
- \frac{ 83262268260174603393926795718797 }%
{ 77633639847061371224064 \, K^{10} }
\nonumber\\[1.377ex]
& \left. - \frac{ 114875486821914130002661264320911491 }%
{ 5589622068988418728132608 \, K^{11} }
+ {\mathcal O}\left( \frac{1}{K^{12}} \right) \right\} \,.
\end{align}
For the ground and the first excited state of the cubic and
quartic potentials, we have put special emphasis
on the evaluation of the tenth-order corrections in order
to illustrate the potential applications of the
methods discussed in the current work.

%
% Results for the Quintic Potential
%
\section{Quintic Potential}
\label{quintic}

The Hamiltonian is used in the normalization:
$h_5(g) \, = \, -\frac{1}{2} \,
\frac{\partial^2}{\partial q^2}  + \frac{1}{2} \, q^2 +  \sqrt{g} \, q^5$,
and the result for $B_5(E,g)$ reads
\begin{align}
\label{B_series_quintic}
& B_5(E, g) = E + g \left( \frac{1107}{256} +
\frac{1085}{32} \, E^2 + \frac{315}{16} \, E^4 \right)
\\[1.377ex]
& \; + g^2 \, \left( \frac{118165905}{8192} \, E +
\frac{96201105}{2048} \, E^3 +
\frac{15570555}{512} \, E^5 +
\frac{692835}{128} \, E^7 \right)
\nonumber\\[1.377ex]
& + \; g^3 \, \left(
\frac{36358712597025}{4194304} +
\frac{142306775756145}{1048576} \, E^2 +
\frac{30926063193025}{131072} \, E^4 \right.
\nonumber\\[1.377ex]
& \; \left. +
\frac{4140194663605}{32768} \, E^6 +
\frac{456782651325}{16384} \, E^8 +
\frac{9704539845}{4096} \, E^{10} \right)
+ \mathcal{O}(g^4) \, .
\nonumber
\end{align}
The perturbative coefficients describing the ground-state energy
diverge rapidly, even in lower orders in $g$,
\begin{align}
\label{P_series_quintic}
\epsilon_0^{(5)}(g) =& \; \frac12 -
\frac{449}{32}\, g -
\frac{1723225}{128}\, g^2 -
\frac{928230645}{16}\, g^3 - \frac{21855598127812155}{32768}\, g^4 +
\mathcal{O}(g^5) \, .
\end{align}
It is of interest to observe that
the leading term of the $A$ function can be
written in terms of a single Gamma function,
although from Eq.~(\ref{instOdd}), one might have
assumed that the result necessarily involves the
expression $B(5/3, 5/3) = \Gamma^2(5/3)/\Gamma(10/3)$.
Throughout the remainder of this article,
we attempt to write the final expressions with the least number
of Gamma functions possible, with the Gamma functions entering
only the numerators,
\begin{equation}
A_5 (E, g) =
\frac{3\,\sqrt{3}\,\,\Gamma^3(\tfrac{2}{3})}{7 \, \pi\,(2 g)^{1/3}} +
{\mathcal O}(g^{1/3} )\,.
\end{equation}
Our new result concerns higher-order terms of the $A$ function.
Perhaps quite surprisingly, Gamma functions enter at orders
$g^{1/3}$ and $g^{2/3}$, but not at order $g$,
\begin{align}
\label{A5}
A_5(E,g) =&  \;
\frac{3\,\sqrt{3}\,\,\Gamma^3(\tfrac{2}{3})}{7 \, \pi\,(2 g)^{1/3}} -
g^{1/3}\, \frac{3 \sqrt{3} \,
\Gamma^3( \tfrac{1}{3} )}{ 2^{2/3} \, 8 \, \pi} \,
\left( \frac{11}{54} + \frac{14}{27} \, E^2 \right)
\nonumber\\[1.377ex]
& \; + g^{2/3}\, \frac{ \Gamma^3(\tfrac{2}{3} ) }{ 2^{1/3} \, \sqrt{3}\, \pi }\,
\left( \frac{385}{32} \, E + \frac{ 935 }{ 72 } \, E^3 \right)
\nonumber\\[1.377ex]
& \; + g\, \left( \frac{21171}{1024}  +
\frac{132245}{1152}\, E^2 +
\frac{10865}{192}\, E^4 \right)
+ {\mathcal O}(g^4) \,.
\end{align}
In view of ${\cal C}(5) = 2^{2/3}$, we have for the quintic potential,
\begin{equation}
\frac{1}{\Gamma\left( \half - B_5(E,g)\right)} =
\frac{1}{\sqrt{8 \pi}}\,
\left( - \frac{ 2^{5/3} }{g^{1/3}} \right)^{B_5(E,g)} \,
\exp(-A_5(E,g)) \,.
\end{equation}
The higher-order corrections for the
decay width of the ground-state resonance energy read as follows,
\begin{align}
\label{im5}
{\rm Im} \, \epsilon_0^{(5)}(g > 0) =&
- \frac{1}{2^{2/3}\,\sqrt{\pi}\,g^{1/6}} \,
\exp\left(
-\frac{\sqrt{3}\,\Gamma^3(\tfrac{1}{3})}{7 \, \pi\,(2 g)^{1/3}} \right)\,
\left( 1 +
\frac{\sqrt{3}\, \Gamma^3(\tfrac{1}{3})}{2^{1/3}\, 8\, \pi} \, g^{1/3} +
\right.
\nonumber\\[1.377ex]
& \left. + \left(
\frac{3 \, \Gamma^6(\tfrac{1}{3})}{2^{1/3}\, 256\,\pi^2} -
\frac{275\,\Gamma^3(\tfrac{2}{3})}{2^{1/3}\, 36\, \sqrt{3} \, \pi}
\right) \, g^{2/3} +
\right.
\nonumber\\[1.377ex]
& \left.+
\left( - \frac{111353}{1152} -
\frac{275\,\pi}{216 \, \sqrt{3}} +
\frac{\sqrt{3} \, \Gamma^9(\tfrac{1}{3})}{4096\, \pi^3}
\right) \, g +
{\mathcal O}(g^{4/3}) \right)
\end{align}
The higher-order corrections for the
factorial asymptotics of the perturbative coefficients thus read,
for the ground state,
\begin{align}
\label{LO5}
& \epsilon_{0,K}^{(5)} = - 3\,\frac{ 7^{3 K+1/2} \, \Gamma(\third)^{9K + 3/2}}%
{ 2^{8K+2} \, \pi^{6 K + 5/2} } \, \Gamma\left(3 K +\half\right)\,
\left\{ 1 + \frac{\pi}{14 \, \sqrt{3}\, K} \right.
\nonumber\\[1.377ex]
& + \frac{1}{K^2} \, \left(
\frac{ \pi }{ 84 \, \sqrt{3} }
+ \frac{ \pi^2 }{ 1176 } -
\frac{ 275\,\Gamma^9(\tfrac{2}{3}) }{ 1176 \, \sqrt{3} \, \pi^3 }
\right)
+ \frac{1}{K^3} \, \left(
\frac{ \pi }{ 504 \, \sqrt{3} } +
\frac{ \pi^2 }{ 1764 } +
\frac{ \pi^3 }{ 49392 \, \sqrt{3} } \right.
\nonumber\\[1.377ex]
& \quad \left. \left. -
\left(
\frac{ 1125377 }{ 790272 \, \sqrt{3}\, \pi^3 } +
\frac{ 275 }{ 49392 \, \pi^2 }
\right) \, \Gamma^9(\tfrac{2}{3}) \right)
+ {\mathcal O}\left( \frac{1}{K^4} \right) \right\} \,.
\end{align}

%
% Results for the Sextic Potential
%
\section{Sextic Potential}
\label{sextic}

With the Hamiltonian in the normalization:
$H_6(g) = -\frac{1}{2} \,
\frac{\partial^2}{\partial q^2}  + \frac{1}{2} \, q^2 + g \, q^6$,
the result for $B_6(E,g)$ reads
\begin{align}
\label{B_series_sextic}
& B_6(E, g) = E - g \left( \frac{25}{8} \, E +
\frac{5}{2} \, E^3 \right)
\nonumber\\[1.377ex]
& + g^2 \, \left( \frac{21777}{256} \, E
+ \frac{5145}{32} \, E^3
+ \frac{693}{16} \, E^5 \right)
\nonumber\\[1.377ex]
& - \, g^3 \, \left( \frac{12746305}{2048} \, E
+ \frac{8703695}{512} \, E^3
+ \frac{1096095 }{128} \, E^5
+ \frac{ 36465 }{32} \, E^7 \right)
+ \mathcal{O}(g^4) \, ,
\end{align}
which gives rise to the following perturbation series
for the ground state,
\begin{equation}
\label{P_series_sextic}
E_0^{(6)}(g) = \frac12 +
\frac{15}{8}\, g -
\frac{3495}{64}\, g^2 +
\frac{1239675}{256}\, g^3 -
\frac{3342323355}{4096}\, g^4 + \mathcal{O}(g^5) \, .
\end{equation}
Our new result concerns higher-order terms of the $A$ function,
\begin{eqnarray}
A_6(E,g) &=&
\frac{ \pi }{2^{5/2} \, (-g)^{1/2}}
- g \, \left( \frac{221}{24} \, E + \frac{17}{3} \, E^3 \right)
\nonumber\\[1.377ex]
&& + g^2 \, \left( \frac{ 2504899 }{ 7680 } \, E + \frac{ 45769 }{ 96 } \, E^3 +
   \frac{ 17527 }{ 160 } \, E^5  \right)
+ {\mathcal O}( g^3 ) \,.
\end{eqnarray}
The special role of the sextic potential is illustrated by the fact
that by analogy with the quintic potential,
one would have expected nonvanishing terms of orders
$g^{1/2}$ and $g^{3/2}$ in the above result.
The surprising lack of these coefficients explains the
result of a lack of a $1/K$-correction to the leading
asymptotic growth of the perturbative coefficients, as is
evident from Eq.~(\ref{LO6}) below.

With ${\cal C}(6) = \sqrt{2}$, the conjectured quantization
condition for the sextic potential reads,
\begin{equation}
\frac{1}{\Gamma\left( \half - B_6(E,g)\right)} =
\frac{1}{\sqrt{2 \pi}}\,
\left( \frac{2^{3/2}}{g^{1/2}} \right)^{B_6(E,g)} \,
\exp(-A_6(E,g)) \,.
\end{equation}
For the ground-state resonance energy at negative coupling,
the following higher-order corrections are easily derived,
\begin{align}
\label{im6}
& {\rm Im} \, E_0^{(6)}(g < 0) =
- \exp\left( -\frac{ \pi }{2^{5/2}\, (-g)^{1/2} } \right)\,
\frac{1}{\sqrt{\pi}}
\nonumber\\[1.377ex]
& \; \quad \times \left( - \frac{2}{g} \right)^{ 1/4 } \,
\left( 1
+ \frac{ 165 }{ 16 }  \, g
- \frac{ 174241 }{ 512 }  \, g^2
+ \frac{ 328657111 }{ 24576 }  \, g^3
+ {\mathcal O}( g^4 ) \right) \,.
\end{align}
Because of the lack of a term
of relative order $\sqrt{g}$ in this expression,
a term of relative order $K^{-1}$ is thus missing from the
corrections to the leading factorial growth of the
perturbative coefficients. The first few correction
terms read for the ground state,
\begin{align}
\label{LO6}
E_{0,K}^{(6)} =&
(-1)^{K+1} \frac{ 2^{5 \, K + 5/2 }}{\pi^{ 2\,K + 2 } }
\Gamma\left( 2 K + \half \right)
\left\{ 1 -  \frac{ 165 \pi^2 }{ 2048 \, K^2} -
\frac{ 165 \pi^2 }{ 2048 \, K^3 }
\right.
\nonumber\\[1.377ex]
& \left. + \frac{1}{K^4} \, \left( - \frac{2145 \pi^2 }{32768 } -
\frac{ 174241 \pi^4  }{ 8388608 } \right) +
{\mathcal O}\left( \frac{1}{K^4} \right) \right\}.
\end{align}

%
% Results for the Septic Potential
%
\section{Septic Potential}
\label{septic}

For the septic potential, we encounter the most complex analytic
expressions so far in our analysis. The
Hamiltonian is used in the normalization
$h_7(g) \, = \, -\frac{1}{2} \,
\frac{\partial^2}{\partial q^2}  + \frac{1}{2} \, q^2 +
\sqrt{g} \, q^7$. The first terms of the
result for $B_7(E,g)$ read as follows,
\begin{align}
\label{B_series_septic}
& B_7(E, g) = E + g \left( \frac{180675}{2048} +
\frac{444381}{512} \, E^2
+ \frac{82005}{128} \, E^4
+ \frac{3003}{32} \, E^6 \right)
\nonumber\\[1.377ex]
& + g^2 \, \left( \frac{182627818702875}{2097152} \, E +
\frac{ 156916927352185 }{  524288 } \, E^3 +
\frac{ 13513312267455  }{ 65536 } \, E^5  \right.
\nonumber\\[1.377ex]
& \qquad \left. +
\frac{ 824707412529 }{ 16384 } \, E^7 +
\frac{ 43689020375 }{ 8192 } \, E^9 +
\frac{ 456326325 }{ 2048 } \, E^{11}  \right)
+ \mathcal{O}(g^3) \, .
\end{align}
The perturbation series for the ground state is a solution of
$B_7(\epsilon_0^{(7)}(g), g) = \half$,
\begin{align}
\label{P_series_septic}
\epsilon_0^{(7)}(g) =& \frac12
- \frac{ 44379 }{ 128 }\, g
- \frac{ 715842493569  }{ 8192 }\, g^2
+ \mathcal{O}(g^3) \, .
\end{align}
The new result for the $A$ function contains
the golden ratio $(\sqrt{5}+1)/2$,
\begin{align}
& A_7(E,g) =
\frac{5^{1/4} \, \Gamma(\tfrac{1}{5} )\,\Gamma(\tfrac{2}{5})}%
  {2^{1/10} \, ( \sqrt{5} + 1 )^{1/2}\,9 \, \pi \, g^{1/5}} +
g^{1/5} \frac{5^{1/4} \,
\Gamma^2( \tfrac{3}{5} )\,\Gamma( \tfrac{4}{5} )}%
{2^{9/10} \, (\sqrt{5} + 1)^{1/2} \, \pi} \,
\left(\frac{5}{8} + \frac{9}{10} \, E^2 \right)
\nonumber\\[1.377ex]
& - g^{2/5}\, \frac{5^{1/4} \, ( \sqrt{5} + 1 )^{1/2} \,
\Gamma^2( \tfrac{1}{5} ) \,
\Gamma( \tfrac{3}{5} ) }{2^{3/10}  \, \pi }\,
\left( \frac{377}{1600} \, E + \frac{299}{2000} \, E^3 \right)
\nonumber\\[1.377ex]
& + g^{3/5} \, \frac{5^{1/4} \, \Gamma( \tfrac{2}{5} ) \,
\Gamma^2( \tfrac{4}{5} ) }{2^{7/10}  \,
( \sqrt{5} - 1 )^{1/2} \, \pi }\,
\left( \frac{59143}{9600} +
\frac{15351}{400} \, E^2 +
\frac{13209}{1000} \, E^4 \right)
+ {\mathcal O}(g^{4/5}) \,.
\end{align}
The quantization condition for $g > 0$, in view of
${\cal C}(7) = 2^{2/5}$, reads
\begin{equation}
\frac{1}{\Gamma\left( \half - B_7(E,g)\right)} =
\frac{1}{\sqrt{8 \pi}}\,
\left( -\frac{2^{7/5}}{g^{1/5}} \right)^{B_7(E,g)} \,
\exp(-A_7(E,g)) \,.
\end{equation}
The analytic result for the decay width of the ground state reads, in
higher order,
\begin{align}
\label{im7}
& {\rm Im} \, \epsilon_0^{(7)}(g > 0) =
- \frac{1}{2^{4/5}\,\sqrt{\pi}\,g^{1/10}} \,
\exp\left(
-\frac{5^{1/4} \, \Gamma(\tfrac{1}{5} )\,\Gamma(\tfrac{2}{5})}%
  {2^{1/10} \, 9 \, \pi \, ( \sqrt{5} + 1 )^{1/2}\,g^{1/5}}
\right)
\nonumber\\[1.377ex]
& \times \left\{ 1 -
g^{1/5} \, \frac{17 \, \Gamma^2(\tfrac{3}{5} ) \,  \Gamma(\tfrac{4}{5} ) }
  {2^{9/10} \, 5^{3/4} ( \sqrt{5} + 1 )^{1/2} \, 4 \pi } +
  g^{2/5} \, \left(
\frac{273 \, ( \sqrt{5} + 1)^{1/2} \,
\Gamma^2( \tfrac{1}{5} ) \, \Gamma( \tfrac{3}{5} )}%
  {2^{3/10} \, 5^{3/4} \, 400 \pi}
\right. \right.
\nonumber\\[1.377ex]
& \left. \left. \quad +
\frac{289 \, \Gamma^4( \tfrac{3}{5} ) \, \Gamma^2( \tfrac{4}{5} )}%
  {2^{4/5} \, 5^{1/2} \, ( \sqrt{5} + 1)^{1/2} \, 320 \pi^2} \right)
+ {\mathcal O}\left(g^{3/5} \right) \right\} \,.
\end{align}
The higher-order asymptotics to the perturbative coefficients are found as
\begin{align}
\label{LO7}
& \epsilon_{0,K}^{(7)} =
- \frac{ 5 \, \Gamma\left(5 K +\half \right)}{ 2^{2K+1} \, \pi^{3/2}} \,
\left( \frac{ 2^{1/2} \, 9 \, ( \sqrt{5} + 1)^{1/2} \, \pi }%
  { 5^{1/4} \, \Gamma(\tfrac{1}{5}) \,
    \Gamma^2(\tfrac{2}{5})} \right)^{5 K + 1/2}
\nonumber\\[1.377ex]
& \left\{ 1 - \frac{2^{1/2}\, 17\, \pi}%
  {5^{1/4}\, (\sqrt{5} + 1)^{3/2} \, 225\, K}
+ \frac{1}{K^2} \, \left(
-\frac{ 17\, \pi }%
  {5^{3/4}\, 2^{1/2}\, (\sqrt{5} + 1)^{1/2} \, 900}
\right. \right.
\nonumber\\[1.377ex]
& \qquad \left. + \frac{ 17\, \pi }%
  {5^{1/4}\, 2^{1/2}\, (\sqrt{5} + 1)^{1/2} \, 4500}
-\frac{ 289\, \pi^2 }%
  {(\sqrt{5} + 1) \, 405000}
\right.
\nonumber\\[1.377ex]
& \qquad \left. \left.
+\frac{ 289\, \pi^2 }%
  {5^{1/2}\, (\sqrt{5} + 1) \, 135000}
+ \frac{ 91\, \Gamma^4(\tfrac{1}{5} ) \,
\Gamma^3(\tfrac{2}{5} ) }%
  {5^{1/2}\, (\sqrt{5} + 1) \, \pi^2 \, 135000}
\right)
+ {\mathcal O}\left( \frac{1}{K^3} \right) \right\} \,.
\end{align}
are found to be rather involved in their analytic structure.

%
% Results for the Octic Potential
%
\section{Octic Potential}
\label{octic}

As we use the Hamiltonian in the
normalization
$H_8(g)  =  -\frac{1}{2} \,
\frac{\partial^2}{\partial q^2}  + \frac{1}{2} \, q^2 + g \, q^8$,
we find the result for $B_8(E,g)$,
\begin{align}
\label{B_series_octic}
& B_8(E, g) = E - g \left( \frac{315}{128} +
\frac{245}{16} \, E^2 + \frac{35}{8} \, E^4 \right)
\nonumber\\[1.377ex]
& + g^2 \, \left( \frac{5604849}{2048} \, E
+ \frac{3209745}{512} \, E^3
+ \frac{291291}{128} \, E^5
+ \frac{6435}{32} \, E^7 \right)
+ \mathcal{O}(g^3) \, .
\end{align}
Solving $B_8(E^{(8)}_0(g), g) = \half$, we find for the ground state:
\begin{equation}
\label{P_series_octic}
E_0^{(8)}(g) = \frac12 +
\frac{105}{16}\, g -
\frac{67515}{32}\, g^2 +
\frac{401548875}{128}\, g^3 -
\frac{25424096867715}{2048}\, g^4 + \mathcal{O}(g^5) \, .
\end{equation}
The result for the subleading terms of the $A$ function is
much less involved as compared to the septic oscillator and reads,
\begin{align}
& A_8(E,g) =
\frac{\sqrt{3}\, \Gamma^3( \tfrac{1}{3} ) }%
{2^{2/3}\,10\,\pi\, (-g)^{1/3}}
+ (-g)^{1/3} \, \frac{ \Gamma^3( \tfrac{2}{3} ) }%
{2^{1/3}\,\sqrt{3}\,\pi} \,
\left( \frac{17}{16} + \frac{5}{4} \, E^2 \right)
\nonumber\\[1.377ex]
& \: - (-g)^{\frac23} \frac{ \Gamma^3( \tfrac{1}{3} ) }%
{2^{\frac23} \sqrt{3} \pi} \,
\left( \frac{77}{96} + \frac{91 E^2}{216} \right)
- g \left( \frac{28007}{2560} +
\frac{22669}{576} E^2 +
\frac{2587}{288} E^4 \right)
+ {\mathcal O}(g^{\frac43}).
\end{align}
The coefficients of the terms of orders $g^{1/3}$ and $g^{2/3}$
involve Gamma functions, whereas the term of order $g$ is a
polynomial in $E$ with rational coefficients, in analogy to the
corresponding result for the quintic oscillator recorded in
Eq.~(\ref{A5}). In view of ${\cal C}(8) = 2^{1/3}$, we have the
following quantization condition for $g < 0$,
\begin{equation}
\frac{1}{\Gamma\left( \half - B_8(E,g)\right)} =
\frac{1}{\sqrt{2 \pi}}\,
\left( \frac{2^{4/3} }{g^{1/3}} \right)^{B_8(E,g)} \,
\exp(-A_8(E,g)) \,.
\end{equation}
By virtue of the dispersion relation (\ref{dispEven}),
the following result for the decay width in higher order,
\begin{align}
\label{im8}
& {\rm Im} \, E_0^{(8)}(g < 0) =
- \exp\left( -
\frac{\sqrt{3}\, \Gamma^3( \tfrac{1}{3} ) }%
{2^{2/3}\,10\,\pi\, (-g)^{1/3}} \right)\,
\frac{1}{\sqrt{\pi}} \,
\left( - \frac{2}{g} \right)^{1/6}
\nonumber\\[1.377ex]
& \left( 1
- (-g)^{1/3} \,
\frac{11 \, \Gamma^3( \tfrac{2}{3} )}{ 2^{1/3}\,8\,\sqrt{3}\,\pi}
+ (-g)^{2/3} \, \left(
\frac{49 \, \Gamma^3(\tfrac{1}{3})}{2^{2/3}\, 108 \, \sqrt{3}\,\pi}
+ \frac{121 \, \Gamma^6(\tfrac{2}{3})}{2^{2/3} \, 384 \, \pi^2 }  \,
\right) \right.
\nonumber\\[1.377ex]
& \left. - (-g) \, \left( \frac{12429}{320}
+ \frac{539 \, \pi }{1944\,\sqrt{3}}  +
\frac{1331 \, \Gamma^9( \tfrac{2}{3} ) }{18432\,\sqrt{3}\, \pi^3} \,
\right)
+ {\mathcal O}(g^{4/3} ) \right) \,,
\end{align}
translates into the following corrections to the leading
factorial growth of the coefficients:
\begin{align}
\label{LO8}
E_{0,K}^{(8)} =& \;
(-1)^{K+1} \, \frac{3}{\pi^{3/2} \, 2^K} \,
\left(
\frac{\Gamma( \tfrac{8}{3} )}{\Gamma( \tfrac{4}{3} )^2} \right)^{3 K + 1/2} \,
\Gamma\left(3 \, K + \half \right) \,
\left\{ 1 -
\frac{ 11 \, \pi }{ 180 \, \sqrt{3} \, K} \right.
\nonumber\\[1.377ex]
& \left. + \frac{1}{K^2} \, \left( - \frac{11 \, \pi }{ 1080 \, \sqrt{3} }
+ \frac{121 \, \pi^2 }{ 194400 }
+ \frac{49 \, \Gamma^9( \tfrac{1}{3} )}{ 129600 \, \sqrt{3} \, \pi^3 }
\right)
+ {\mathcal O}\left( \frac{1}{K^3} \right) \right\} \,.
\end{align}

\begin{table*}
\begin{scriptsize}
\caption{\label{table1} General properties of even and odd
anharmonic oscillators of arbitrary degree (even integer $N$
and odd integer $M$). We summarize the most important
interconnections between different regions in the
complex coupling parameter plane.}
\begin{center}
\begin{tabular}{cccc}
\hline
\hline
\rule[-2mm]{0mm}{6mm}
Odd, Unstable & Even, Unstable & Odd, \PT{} (Stable) & Even, Stable \\
\rule[-2mm]{0mm}{6mm}
(positive coupling) & (negative coupling) &
(negative coupling) & (positive coupling) \\
\hline
\hline
\rule[-2mm]{0mm}{8mm}
$h_M(g) = {\displaystyle -\frac12 \, \frac{\partial^2}{\partial q^2}}$ &
$H_N(-g) = {\displaystyle -\frac12 \, \frac{\partial^2}{\partial q^2}}$ &
$h_M(-g) = {\displaystyle -\frac12 \, \frac{\partial^2}{\partial q^2}}$ &
$H_N(g) = {\displaystyle -\frac12 \, \frac{\partial^2}{\partial q^2}}$ \\
\rule[-4mm]{0mm}{8mm}
${\displaystyle + \frac12\, q^2 \pm |\sqrt{g}| \, q^M}$ &
${\displaystyle + \frac12\, q^2 - |g| \, q^N}$ &
${\displaystyle + \frac12\, q^2 \pm \ii |\sqrt{g}| \, q^M}$ &
${\displaystyle + \frac12\, q^2 + |g| \, q^N}$ \\
\hline
\rule[-2mm]{0mm}{6mm}
Under complex scaling: &
Under complex scaling: &
No resonances &
No resonances \\
\rule[-2mm]{0mm}{6mm}
Resonances appear &
Resonances appear &
&
\\
\hline
\rule[-2mm]{0mm}{6mm}
Perturbation theory is &
Perturbation theory is &
Perturbation theory is &
Perturbation theory is \\
\rule[-2mm]{0mm}{6mm}
not Borel summable &
not Borel summable &
Borel--Leroy summable &
Borel--Leroy summable \\
\rule[-2mm]{0mm}{6mm}
in the ordinary sense, &
in the ordinary sense, &
to the real eigenvalue &
to the real eigenvalue \\
\rule[-2mm]{0mm}{6mm}
but Borel--Leroy summable &
but Borel--Leroy summable &
&
\\
\rule[-2mm]{0mm}{6mm}
in the distributional sense &
in the distributional sense &
&
\\
\hline
\rule[-2mm]{0mm}{7mm}
\includegraphics[width=1.0cm,height=0.5cm]{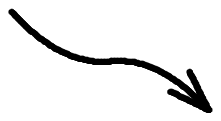} &
\includegraphics[width=1.0cm,height=0.5cm]{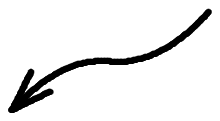} &
\includegraphics[width=1.0cm,height=0.5cm]{arrow1.eps} &
\includegraphics[width=1.0cm,height=0.5cm]{arrow2.eps} \\
\multicolumn{2}{c}{
\rule[-5mm]{0mm}{10mm}
$\displaystyle{
\frac{\Gamma\left( \half - B_M(E, g)\right)}%
{\sqrt{8 \pi} \,\, \ee^{A_M(E, g)}}\,
\left( - \frac{2 \, {\cal C}(M)}{g^{1/(M-2)}} \right)^{B_M(E, g)}} = 1$} &
\multicolumn{2}{c}{
$\displaystyle{
\frac{1}{\Gamma\left( \half - B_M(E,g)\right)} = 0 }$} \\
\multicolumn{2}{c}{
\rule[-5mm]{0mm}{10mm}
$\displaystyle{
\frac{\Gamma\left( \half - B_N(E,g)\right)}%
{\sqrt{2 \pi} \,\, \ee^{A_N(E,g)}}\,
\left(-\frac{2 \, {\cal C}(N)}{(-g)^{2/(N-2)}} \right)^{B_N(E,g)}} = 1$} &
\multicolumn{2}{c}{
$\displaystyle{
\frac{1}{\Gamma\left( \half - B_N(E,g)\right)} = 0 }$} \\
\hline
\rule[-2mm]{0mm}{10mm}
\includegraphics[width=1.0cm,height=0.5cm]{arrow1.eps} & 
[Unstable/stable Odd Oscillators] &
\includegraphics[width=1.0cm,height=0.5cm]{arrow2.eps} & \\
\multicolumn{3}{c}{
\rule[-6mm]{0mm}{14mm}
$\displaystyle{\epsilon^{(M)}_n(g) = n + \frac12 +
\frac{g}{\pi}\, \int_0^{\infty} {\mathrm d} s \,
\frac{{\rm Im} \, \epsilon^{(M)}_n(s + \ii\,0)}{s\, (s - g)}}$} & \\
\multicolumn{3}{c}{
\rule[-6mm]{0mm}{14mm}
$\displaystyle{\epsilon^{(M)}_n(g) \sim
\sum_K \epsilon^{(M)}_{n,K} \, g^K}$} & \\
\multicolumn{4}{l}{
\rule[-6mm]{0mm}{14mm}
$\displaystyle{
\epsilon_{n, K}^{(M)} \sim -\frac{M-2}{\pi^{3/2} \, n! \, 2^{2 K + 1 - n}} \,
\left[ B\left(\frac{M}{M-2}, \frac{M}{M-2}\right) \right]^{-(M-2)\, K - n - 1/2} \,
\Gamma\left[ (M-2)\, K + n + \frac12 \right] }$} \\
\hline
\rule[-2mm]{0mm}{10mm}
& \includegraphics[width=1.0cm,height=0.5cm]{arrow1.eps} & 
[Unstable/stable Even Oscillators] &
% & \includegraphics[width=1.0cm,height=0.5cm]{arrow1.pdf} & &
\includegraphics[width=1.0cm,height=0.5cm]{arrow2.eps} \\
& \multicolumn{3}{c}{
$\displaystyle{E^{(N)}_n(g) = n + \frac12 - \frac{g}{\pi}\, \int_{-\infty}^0
{\mathrm d} s \,
\frac{{\rm{Im}}\, E^{(N)}_n(s+{\rm i}\,0)}{s \, (s - g)}}$} \\
& \multicolumn{3}{c}{
\rule[-6mm]{0mm}{14mm}
$\displaystyle{E^{(N)}_n(g) \sim \sum_K E^{(N)}_{n,K} \, g^K}$} \\
\multicolumn{4}{r}{
\rule[-6mm]{0mm}{14mm}
$\displaystyle{
E_{n, K}^{(N)} \sim \frac{(-1)^{K+1} (N-2)}{\pi^{3/2} \,  n! \, 2^{K + 1 - n}} \,
\left[ B\left(\frac{N}{N-2}, \frac{N}{N-2}\right) \right]^{- (N - 2) \, K/2 - n - 1/2 }
\, \Gamma\left( \frac{N-2}{2} \, K + n + \frac12 \right) }$} \\
\hline
\hline
\end{tabular}
\end{center}
\end{scriptsize}
\end{table*}

%
% General Properties of Even versus Odd Oscillators
%
\chapter{Views from Different Perspectives}
\label{view}

\vspace*{-0.4cm}
\textcolor{light}{ \rule{\textwidth}{0.2cm} }

%
% General Properties of Even versus Odd Oscillators
%
\section{Even and Odd Oscillators: An Overview}
\label{master}

In order to view the investigations reported here from a broader perspective,
let us remember here that originally, the
investigations on anharmonic oscillators started with the
consideration of the cubic and quartic cases as paradigmatic
examples for the lowest-order odd and even anharmonic oscillators.
One can the generalize these in many ways: to an internal symmetry
$O(N)$~\cite{BrLGZJ1977prd1}, to more space-time dimensions
(toward field theory, see~\cite{Ba1960,MK1979}),
and to anharmonic terms of arbitrary degree~\cite{BeWu1971,BeWu1973}.
The latter generalization is the one studied by us here.
In the broader context of the various generalizations mentioned,
the dispersion relations (\ref{dispOdd}) and~(\ref{dispEven})
might be of particular interest~\cite{BeDu1999}. It has often been asked whether
cubic theories and those with higher-order odd
perturbations represent physically self-consistent
theoretical models. Finally, their Hamiltonians are
not bounded from below for field configurations which accumulate
near the region where the potential assumes large negative values.
Based on the considerations presented here,
we can say that if one continues
the coupling into the complex plane, the ``odd'' theories are
no less natural than the ``even'' ones: namely, the
``stable'' regions for the coupling parameter $g$ (the domain
of positive $g$ and the domain of purely imaginary $g$,
respectively) are connected to the ``unstable''
regions (negative $g$ for the ``even'' theories and positive
$g$ for the ``odd'' theories) via dispersion relations.

One can ask the question why we construct the dispersion
relations (\ref{dispOdd}) and (\ref{dispEven}) with
resonances, i.e.~by assigning to the ``unstable'' parameter
regions resonance eigenvalues with a nonvanishing
imaginary part. The corresponding eigenfunctions
fulfill complex-rotated boundary conditions~\cite{BaCo1971}.
Finally, one can also associate
$L^2(\mathbbm{R})$-boundary conditions
to the quartic potential, which leads to a spectrum that is
unbounded from below but discrete (see~\cite{FePe1995}
for an instructive discussion).
The reason is that every real energy eigenvalue has
a natural analytic continuation to the
complex coupling plane only when the boundary
conditions are rotated, i.e.~in that case it is possible to
find a spectrum of resonances whose real parts are in fact
bounded from below, even for the ``unstable'' parameter regions.
As stressed in~\cite{BeDu1999},
the dispersion relations associate the Herglotz
properties of the resonance eigenvalues
with some global analytic properties in the cut complex
coupling plane.

The most important properties of the
anharmonic oscillators studied here are summarized in Table~\ref{table1}.
A few literature references regarding the entries are in order.
Recent numerical investigations regarding the
resonances in the cubic and quartic potentials, for the
the unstable parameter regions,
can be found in~\cite{KiGrJo2004,KiGrJo2006,JeSuLuZJ2008},
and important steps toward a strong-coupling analysis of the
resonance eigenvalues of the cubic oscillator have been
published in~\cite{SeGo1998,FeGuRoZn2005}.
For the cubic, \PT{}-symmetric Hamiltonian,
the boundedness of the spectrum (from below) and the discreteness
have been proven in~\cite{DoDuTa2002i,DoDuTa2002ii}.
According to~\cite{BeWe2001}, there is numerical
evidence that the perturbation
series for the cubic, \PT{}-symmetric Hamiltonian
is Stieltjes, and
that the results of Pad\'{e} prediction techniques
are indistinguishable from those for the quartic oscillator,
which is known to be Borel-summable.
Note that the Stieltjes property alone would already imply Borel
summability.  We would like to conjecture here a generalization of this
property to all the odd, \PT{}-symmetric
anharmonic oscillators: namely, that
their perturbation series are Borel summable to the energy eigenvalue.
This conjecture is inspired by the fact that recently~\cite{Ca2000},
the distributional Borel summability~\cite{CaGrMa1986} has been proven for
odd, unstable anharmonic oscillator of an arbitrary degree.
We also recall that the dispersion relation joining the
odd, unstable oscillators with the odd, stable oscillators
has been given in~\cite{BeDu1999}, whereas the
corresponding relation for the even oscillators
has been derived much earlier~\cite{LoMaSiWi1969,BeWu1971,BeWu1973}.
We also recall the proof of the Borel summability of the
perturbation series for the quartic oscillator~\cite{GrGrSi1970}.
The necessary generalization of Borel summability to Borel--Leroy
summability for oscillators of higher degree has been
discussed recently~\cite{CaEtAl2007}.

\begin{figure}[tb!]
\begin{center}
\begin{minipage}{0.7\linewidth}
\begin{center}
\includegraphics[width=0.7\linewidth]{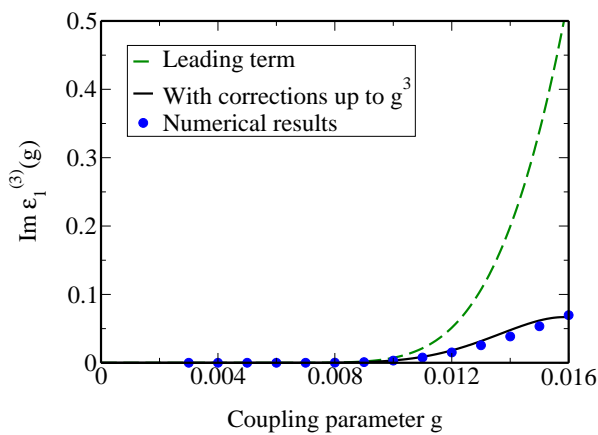}
\caption{\label{fig2} Comparison of the analytic
results (\ref{im3level1recall}) and
(\ref{im3level1leading}) with numerical data for the
imaginary part ${\rm Im} \, \epsilon_1^{(3)}(g)$
of the resonance energy of the first excited state of
the cubic potential.}
\end{center}
\end{minipage}
\end{center}
\end{figure}

%
% Numerical Calculations for Weak Coupling
%
\section{Numerical Calculations for Weak Coupling}
\label{numdiag}

Let us note first of all that all analytic results derived here
have been subjected to extensive numerical
verification. Here, we would only like to
discuss one particular case, the higher-order
formula for the subleading corrections to the
decay width of the first excited level of the
cubic potential, which we recall here from Eq.~(\ref{im3level1}),
\begin{align}
\label{im3level1recall}
& {\rm Im} \, \epsilon_1^{(3)}(g) =
- \frac{ 8 \, {\rm e}^{-2/(15\,g)} }{ \sqrt{\pi} \, g^{3/2}} \,
\left\{ 1 - \frac{853}{16} \, g +
\frac{33349}{512} \, g^2 - \frac{395368511}{40960} \, g^3 +
{\mathcal O}(g^4) \right\} \,.
\end{align}
Our purpose is to compare this result to the numerical values
generated exclusively by the leading-order term
\begin{equation}
\label{im3level1leading}
{\rm Im} \, \epsilon_{1,{\rm leading}}^{(3)}(g) =
- \frac{ 8 \, {\rm e}^{-2/(15\,g)} }{ \sqrt{\pi} \, g^{3/2}} \, .
\end{equation}
The relative correction of order $g$ as given by the
term $-853 \,g/16 \approx -53 \, g$ in Eq.~\eqref{im3level1recall}
is extremely important in reaching satisfactory agreement
of the analytic result and numerical data for the imaginary
part of the resonance energy ${\rm Im} \, \epsilon_1^{(3)}(g)$,
even at relatively small coupling parameter $g \leq 0.01$.
This is evident from Fig.~\ref{fig2}, where we compare the
leading term of the asymptotics given in Eq.~\eqref{im3level1leading}
as well as the first three correction terms given in
Eq.~\eqref{im3level1recall} to numerical data.

%
% Numerical Calculations for Large Coupling
%
\section{Numerical Calculations for Large Coupling}
\label{numlarge}

As we look at the dispersion relations (\ref{dispEven}) and (\ref{dispOdd}),
we might ask ourselves how far the cut extends: from $g=0$ to $g=\infty$,
or possibly up to a finite value of the coupling.
Our conjecture here is that the resonances of the
unstable oscillators persist for any value of the coupling,
however large. Intuition might suggest otherwise.
The unstable case is naturally associated with tunneling
through the potential barrier.
As the coupling increases, the barrier becomes narrower
until, in the limit of very large modulus of the coupling, it eventually
disappears. One might assume that only a limited number
of resonances persist for a large, but finite value of the
coupling: namely, only those for which the real part
of the energy lies significantly ``under the top'' of the
barrier. Numerical evidence~\cite{JeSuLuZJ2008} suggests otherwise.
Namely, we investigate the
strong-coupling expansion for the resonances of odd oscillators.
After a scaling transformation $q \to g^{-1/(2 (M+2))} \, q$,
which leaves the spectrum invariant,
the odd Hamiltonian (\ref{hM}) reads
\begin{equation}
\label{scaling_odd}
h_{{\rm s}, M}(g) = g^{1/(M+2)} \,
\left( h_{\ell, M} + \frac12 \, g^{-2/(M+2)} \, q^2 \right) \,,
\qquad 
h_{\ell, M} =  - \frac12 \, \frac{\partial^2}{\partial q^2} + q^M \, ,
\end{equation}
where $h_{\ell, M}$ does not depend on the coupling $g$.
This Hamiltonian now gives the strong-coupling
expansion in a natural way: One can regard the
term $\tfrac{1}{2} \, g^{-2/(M+2)} \, q^2$
as a perturbation of the term $q^M$.
Via a complex scaling transformation $q \to q \, \ee^{\ii \, \theta_M}$,
we transform $h_{\ell, M} \to h_{\ell, M}(\theta)$ where
\begin{align}
\label{rotationOdd}
h_{\ell, M}(\theta) =& \; \ee^{-2 \, \ii \theta_M} \,
\left( - \frac12 \, \frac{\partial^2}{\partial q^2} +
q^M \, \ee^{\ii \, (M + 2) \, \theta_M} \right) \,,
\qquad 
\theta_M = \frac{\pi}{M + 2} \,.
\end{align}
A diagonalization of this complex-scaled Hamiltonian
in the space of harmonic oscillator eigenfunctions gives
us approximations to the resonance energies.
The resonance eigenvalues have a well-defined complex argument,
and they can be numbered by an index $n = 0,1,2,\dots$. They read
\begin{equation}
\epsilon^{(M)}_{\ell,n} =
\left| \epsilon^{(M)}_{\ell,n} \right| \, \ee^{-\ii \, \pi/(M+2)} \,,
\end{equation}
where by virtue of (\ref{dispOdd}), we associate
the resonance as opposed to the antiresonance with the
region infinitesimally above the real axis,
\begin{equation}
\label{sc_odd}
\epsilon^{(M)}_n(g + \ii 0) =
g^{1/(M+2)} \, \left[ \epsilon^{(M)}_{\ell,n} +
{\mathcal O}\left(g^{-2/(M+2)}\right) \right]  \,,
\end{equation}
for $g \to \infty$.
These asymptotics suggest that all resonances $n = 0,1,2, \dots$
persist for any modulus of the coupling.
Specifically, according to Table~2 of~\cite{JeSuLuZJ2008},
we have
\begin{equation}
\epsilon^{(3)}_{\ell,0} = 0.762\,851\,775 \, \ee^{-\ii \, \pi/5} \,,\quad
\epsilon^{(3)}_{\ell,1} = 2.711\,079\,923 \, \ee^{-\ii \, \pi/5} \,,\quad
\epsilon^{(3)}_{\ell,2} = 4.989\,240\,088 \, \ee^{-\ii \, \pi/5} \,.
\end{equation}
The above results
define the large-coupling coupling asymptotics for the
for the first three resonances
$\epsilon^{(3)}_n(g) \to g^{1/5} \, \epsilon^{(3)}_{\ell,n} $,
for $g \to \infty +\ii 0$ (with $n=0,1,2$).
We have also obtained corresponding results for higher
excited states and conjecture here that all resonances
of the cubic oscillator persist for any value of the coupling,
however large. The corresponding large-coupling asymptotics
for the quantic potential are given by the levels
\begin{equation}
\epsilon^{(5)}_{\ell,0} = 0.709\,935\,516 \, \ee^{-\ii \, \pi/7} \,, \quad
\epsilon^{(5)}_{\ell,1} = 2.659\,756\,364 \, \ee^{-\ii \, \pi/7} \,, \quad
\epsilon^{(5)}_{\ell,2} = 5.458\,235\,423\, \ee^{-\ii \, \pi/7} \,.
\end{equation}

For the even oscillators, the corresponding formulas read
as follows.
We recall that the resonances appear for negative $g$,
and that the resonance as opposed to the antiresonance
is attached to the negative $g$ axis for infinitesimal positive
imaginary part of the coupling.
Under a scaling transformation $q \to (-g)^{-1/(N+2)} \, q$,
the Hamiltonian (\ref{HN}) transforms to  the following
scaled Hamiltonian with the same spectrum,
\begin{equation}
\label{scaling_even}
H_{{\rm s}, N}(g) = (-g)^{2/(N + 2)} \,
\left( H_{\ell, N} + \frac12 \, (-g)^{-4/(N+2)} \, q^2 \right) \,, \qquad
H_{\ell, N} = - \frac12 \, \frac{\partial^2}{\partial q^2} - q^N \,.
\end{equation}
As a WKB--approximation to the wave function shows,
the appropriate scaling transformation for even $N$ is
also of the form $q \to q \ee^{\ii \, \theta_N}$, but with
$H_{\ell, N}(\theta)$ and $\theta_N$ given by
\begin{equation}
\label{rotationEven}
H_{\ell, N}(\theta) = \ee^{-2 \ii \theta_N} \,
\left( - \frac12 \, \frac{\partial^2}{\partial q^2} -
q^N \, \ee^{\ii \, (N + 2) \, \theta_N} \right) \,, \qquad
\theta_N = \frac{2 \, \pi}{N + 2} \,.
\end{equation}
The eigenvalues again have a well-defined complex phase,
\begin{equation}
E^{(N)}_{\ell,n} =
\left| E^{(N)}_{\ell,n} \right| \, \ee^{-2 \, \ii \, \pi/(N+2)} \,.
\end{equation}
The large-coupling asymptotics for the resonances then reads
\begin{equation}
\label{sc_even}
E^{(N)}_n(g + \ii 0) = (-g)^{2/(N+2)} \,
\left[ E^{(N)}_{\ell,n} +
{\mathcal O}\left(g^{-4/(N+2)}\right) \right] \,,
\end{equation}
for $g \to -\infty$.
A numerical determination  of the complex resonance
eigenvalues $E^{(N)}_{\ell,n}$ of the non--Hermitian Hamiltonian
$H_{\ell, N}(\theta)$ is thus sufficient in order
determine the large-coupling asymptotics of the
resonance eigenvalues of $H_{{\rm s}, N}(g)$, which
coincide with the resonance eigenvalues
$E^{(N)}_n(g + \ii 0)$ of $H_N(g)$ attained for $g < 0$.
For the first three resonances of the quartic potential,
we find by numerical calculations,
\begin{equation}
E^{(4)}_{\ell,0} = 0.667\,986\,260 \, \ee^{-\ii \, \pi/3} \,, \quad
E^{(4)}_{\ell,1} = 2.393\,644\,016 \, \ee^{-\ii \, \pi/3} \,, \quad
E^{(4)}_{\ell,2} = 4.696\,795\,386\, \ee^{-\ii \, \pi/3} \,.
\end{equation}
This example, which demonstrates the
persistence of the quartic resonances in the
strong-coupling limit, concludes our analysis of the
complex resonances.
Note that the strong-coupling expansions, whose
leading terms are indicated in Eqs.~(\ref{sc_odd})
and~(\ref{sc_even}), are in some sense complementary to the
weak-coupling expansions (\ref{genEvenHigher})
and (\ref{genOddHigher}).

%
% Conclusions
%
\chapter{Conclusions}
\label{conclu}

\vspace*{-0.4cm}
\textcolor{light}{ \rule{\textwidth}{0.2cm} }

Anharmonic oscillators form a very basic model for many physical
processes. The anharmonic terms induce coupling among the
energy levels of the harmonic oscillator, which can be interpreted
as inducing some sort of ``higher-harmonic generation'' due to
the intertwining of an infinite number of harmonic modes in the
energy levels of the anharmonic oscillator. At the same time,
the anharmonic oscillators represent classically integrable systems.
There is no chaos, and therefore, the corresponding
quantum systems are amenable to a unified treatment which
is based on semiclassical expansions and which is manifest
in the generalized quantization conditions
(\ref{quantEven}) and (\ref{quantOdd}).
These quantization conditions allow us to determine
the subleading corrections to the decay widths of
arbitrary levels of anharmonic oscillators of arbitrary degree.

The formulas for the subleading corrections to the
decay width of the resonances, as presented in
Eqs.~(\ref{im3level1}),~(\ref{im4}),
(\ref{im4level1}),~(\ref{im5}), (\ref{im6}),~(\ref{im7})~and~(\ref{im8})
have not yet appeared in the literature to the best of our
knowledge. We note, in particular, that the results for
the excited states in Eqs.~(\ref{im3level1}) and (\ref{im4level1})
are derived on the basis of the generalized quantization conditions;
these could not have been obtained by standard field-theoretical
investigations which involve a higher-order perturbative
expansion about the instanton configurations, as it is
the case for the ground-state results (see also Sec.~\ref{numdiag}).
All analytic results presented here have been subjected to
intensive numerical checks, including those for the
sextic anharmonic oscillator where the higher-order
corrections are observed to have a rather surprising
structure.

Let us now attempt to interpret the results from a broader perspective.
In the case of a theory with an ``unstable'' potential where, classically,
the motion of the particle is not confined to a finite spatial
region, one typically encounters quantum mechanical resonances
with a nonvanishing decay width, when one endows the problem
with ``outgoing'' boundary conditions (and the corresponding
antiresonances for the complex conjugated boundary conditions).
The complex boundary conditions
lead to a non-Hermitian Hamilton operator whose resonance energies
are not real. Intuition suggests that there must be way of
smoothly changing the coupling constant and the boundary
conditions from the ``resonance case'' to the corresponding
complex conjugated antiresonance, so that the resonance
eigenvalue ``transits'' through the real
axis. At this point, precisely, at least for the odd anharmonic oscillators,
the eigenenergies become real, and the Hamiltonian becomes
${\mathcal{PT}}$-symmetric (see~\cite{BeBo1998}).
This is the observation that leads, naturally, to the
dispersion relation \eqref{dispOdd} for odd anharmonic
oscillators.

Analogously~\cite{LoMaSiWi1969,BeWu1971,BeWu1973}, for
even anharmonic oscillators, if one starts from negative coupling, there is a
way of changing the complex argument of the coupling parameter so that the
boundary conditions for resonances and antiresonances deform smoothly into each
other, crossing an entirely real eigenvalue for the energy at positive, and
real, coupling $g$, and this corresponds to the relation \eqref{dispEven}.

In some sense, ${\mathcal{PT}}$-symmetric odd anharmonic oscillators receive an
interpretation as the analogy complete analogy of even anharmonic oscillators
for positive and real coupling.  They provide a bridge between resonances and
antiresonances, as exemplified by the dispersion relation~\eqref{dispOdd},
over the real
axis. They also facilitate the interpretation of theories which allow for the
presence of complex resonance and antiresonance energies, by allowing us to
interpret these theories as the unstable realizations of a stable,
${\mathcal{PT}}$-symmetric theory, continued into the complex plane.

By a similarity transformation~\cite{BeBo1998,Mo2002i,Mo2002iii,Mo2005cubic},
it is possible to map the
dynamics induced by a ${\mathcal{PT}}$-symmetric theory onto a Hermitian
theory, but the resulting expressions can be very complicated, and even in the
case of the relatively simple cubic ${\mathcal{PT}}$-symmetric oscillator, the
corresponding Hermitian Hamiltonian obtained by the similarity transformation
cannot be written down in closed form and contains operators that mix momenta
and coordinates. In that case, the original, ${\mathcal{PT}}$-symmetric theory,
namely the ${\mathcal{PT}}$-symmetric cubic oscillator, provides for a much
more compact formulation that allows for a workable, and practically feasible
formulation in terms of variational principles and corresponding classical
trajectories of the dynamical variables, which would be much more difficult
to achieve when starting the discussion from the much more involved expressions
that occur in a typical case~\cite{Mo2005cubic}
for the equivalent, Hermitian theory.
This is one of the reasons why \PT-symmetric theories have received
some attention in the physics community in recent years.

Although a high degree of unification can been achieved
for odd versus even anharmonic oscillators using concepts
from \PT-symmetry and generalized quantization conditions,
let us finally, that there is some room left for future investigations.
E.g., one might well ask
what the quantization conditions \eqref{dispOdd}
and~\eqref{dispEven} should look like for arbitrary
complex argument of the coupling $g$.
Our conjectures as presented in
Eqs.~(\ref{quantEven}),~(\ref{quantOdd}) and~(\ref{quantStable})
are relevant only for either positive or negative, or purely
imaginary coupling parameter.
These parameter ranges represent the physically
most important scenarios by far, but we would like to
point out that  the interpolation required to
describe an arbitrary complex argument of the coupling parameter
remains an open problem.

Finally, in view of the phenomenological significance of
anharmonic oscillators, we believe that the numerically large
correction terms in the higher-order formulas for the decay widths,
as obtained, e.g.,~in Eqs.~(\ref{im3level1}) and (\ref{im4level1}),
might find interesting applications in the various branches of
the theory of finite quantum systems.

\chapter*{Acknowledgments}

The authors acknowledge helpful discussion with M.~Lubasch.
U.D.J.~acknowledges support by a Grant from the
Missouri Research Board and
by the National Science Foundation (Grant PHY--8555454).
A.S. acknowledges support from the Helmholtz Gemeinschaft
(Nachwuchsgruppe VH--NG--421).

\end{document}